\documentclass[aps,prb,a4paper,twocolumn,10pt,floatfix,showpacs,showkeys,amsmath,amssymb,nobibnotes,unsortedaddress,superscriptaddress]{revtex4-2}

\setcitestyle{super}
\usepackage{graphicx}
\usepackage[export]{adjustbox}
\usepackage[dvipsnames]{xcolor}
\usepackage[utf8]{inputenc}
\usepackage[T1]{fontenc}
\usepackage{amsmath}
\usepackage{xr}
\usepackage{hyperref}
\usepackage{upgreek}
\usepackage[percent]{overpic}
\usepackage{xcolor}
\usepackage{physics}
\usepackage{verbatim}
\usepackage{csquotes}
\usepackage[explicit]{titlesec}
\usepackage{lipsum}
\usepackage{xspace}
\usepackage{lineno}
\usepackage{appendix}
\usepackage{diagbox}
\usepackage{subcaption}
\usepackage{tabularx}
\usepackage{array}
\usepackage{multirow}

\graphicspath{{figures/}}
\renewcommand{\figurename}{Figure}
\renewcommand{\thefigure}{\arabic{figure}}
\renewcommand{\tablename}{Table}
\renewcommand{\thetable}{\arabic{table}}
\renewcommand{\thesection}{\arabic{section}}

\titleformat{\section}{\large\bfseries\filcenter}{\thesection.}{1em}{#1}

\newcommand{\jgen}{J_\mathrm{gen}}

\newcommand{\jsc}{J_\mathrm{sc}}

\newcommand{\Voc}{V_\mathrm{oc}}
\newcommand{\Vocsq}{V_\mathrm{oc}^{\mathrm{SQ}}}
\newcommand{\Vocrad}{V_\mathrm{oc}^{\mathrm{rad}}}
\newcommand{\dVocsq}{\Delta V_\mathrm{oc}^{\mathrm{SQ}}}
\newcommand{\dVocabs}{\Delta V_\mathrm{oc}^{\mathrm{abs}}}
\newcommand{\dVocnr}{\Delta V_\mathrm{oc}^{\mathrm{nonrad}}}
\newcommand{\Vimp}{V_\mathrm{imp}}
\newcommand{\Vext}{V_\mathrm{ext}}

\newcommand{\uoc}{v_\mathrm{oc}}
\newcommand{\FF}{F\!F}

\newcommand{\Rtr}{R_\mathrm{tr}}
\newcommand{\Rext}{R_\mathrm{ext}}
\newcommand{\nid}{n_\mathrm{id}}
\newcommand{\nsig}{n_\mathrm{\sigma}}

\newcommand{\Eg}{E_\mathrm{g}}

\newcommand{\sigmaoc}{\sigma_\mathrm{oc}}

\newcommand{\etaexc}{\eta_\mathrm{exc}}
\newcommand{\EQE}{\mathrm{EQE}_{\mathrm{PV}}}
\newcommand{\bl}{\left(}
\newcommand{\bL}{\left[}
\newcommand{\br}{\right)}
\newcommand{\bR}{\right]}

\newcommand{\kT}{k_B T}
\newcommand{\der}{\mathrm{d}}

\begin{document}

\title{Bridging the lab-to-fab gap in non-fullerene organic solar cells via gravure printing}

\author{Svitlana Taranenko}
\affiliation{Institute for Print and Media Technology, Chemnitz University of Technology, 09126 Chemnitz, Germany}

\author{Chen Wang}
\affiliation{Institute of Physics, Chemnitz University of Technology, 09126 Chemnitz, Germany}

\author{David Holzner}
\affiliation{Institute for Print and Media Technology, Chemnitz University of Technology, 09126 Chemnitz, Germany}

\author{Robert Eland}
\affiliation{Institute for Print and Media Technology, Chemnitz University of Technology, 09126 Chemnitz, Germany}

\author{Christopher Wöpke}
\affiliation{Institute of Physics, Chemnitz University of Technology, 09126 Chemnitz, Germany}

\author{Toni Seiler}
\affiliation{Institute of Physics, Chemnitz University of Technology, 09126 Chemnitz, Germany}

\author{Alexander Ehm}
\affiliation{Institute of Physics, Chemnitz University of Technology, 09126 Chemnitz, Germany}

\author{Fabio Le Piane}
\affiliation{Institute of Physics, Chemnitz University of Technology, 09126 Chemnitz, Germany}

\author{Roderick C.\ I.\ Mackenzie}
\affiliation{Department of Engineering, Durham University, Lower Mount Joy, South Road, Durham DH1 3LE, United Kingdom}

\author{Dietrich R.\ T.\ Zahn}
\affiliation{Institute of Physics, Chemnitz University of Technology, 09126 Chemnitz, Germany}

\author{Carsten Deibel}
\affiliation{Institute of Physics, Chemnitz University of Technology, 09126 Chemnitz, Germany}

\author{Arved Carl Hübler}
\affiliation{Institute for Print and Media Technology, Chemnitz University of Technology, 09126 Chemnitz, Germany}

\author{Maria Saladina}
\email[Corresponding author. Email: ]{maria.saladina@physik.tu-chemnitz.de}
\affiliation{Institute of Physics, Chemnitz University of Technology, 09126 Chemnitz, Germany}

\begin{abstract}

Organic solar cells have reached record efficiencies with non-fullerene acceptors, yet their translation to industrial printing remains a critical bottleneck. Here we report the highest efficiency achieved for a fully roll-to-roll-compatible gravure-printed non-fullerene organic solar cell. High-performance blends are typically optimised under laboratory coating conditions, while roll-to-roll manufacturing imposes fundamentally different constraints on ink stability, drying dynamics, and multilayer integration. Whether these constraints intrinsically limit device physics has remained unresolved. Here, we demonstrate a gravure-printed PM6:Y12 solar cell architecture using commercially available materials and establish a quantitative framework that disentangles optical, recombination, and transport losses in printed devices. We find that favourable bulk morphology and exciton harvesting can be preserved under gravure printing and non-halogenated solvents. The dominant efficiency penalties arise instead from optical interference within the printed layer stack and slow charge transport. Our results demonstrate that the performance gap between laboratory and printed solar cells is originating from device architecture rather than the intrinsic physics of modern non-fullerene systems, providing a mechanistic roadmap for roll-to-roll manufacturing of non-fullerene solar cells.

\end{abstract}


\maketitle

\section{Introduction}

Early demonstrations of printed organic solar cells (OSCs), largely based on fullerene acceptors, established the feasibility of roll-to-roll (R2R) processing more than a decade ago.\cite{hoth2008printing,ding2009patternable,kopola2010high,huebler2011printed,yang2013organic,eggenhuisen2015high,apilo2015roll,kapnopoulos2016fully} With the emergence of non-fullerene acceptors (NFAs), OSC efficiencies have rapidly increased, and lab-scale devices now exceed 20\,\%.\cite{fu2025twostep,jiang2025efficiency,deng2025acceptor} Scalable coating methods such as blade and slot-die coating have shown that high efficiencies can be maintained beyond spin coating.\cite{basu2024large,du2025efficient,feroze2025long} In contrast, true printing of high-performance NFA active layers compatible with continuous R2R manufacturing remains largely unexplored.

Even in the limited reports on printed NFA-based active layers, device fabrication has remained only partially printed.\cite{yang2020large,perkhun2021high,liu2024influence,yang2025high,qi2025roll} Printing -- whether by inkjet or gravure -- is typically combined with spin coating, blade coating, or thermal evaporation for at least one functional layer. This reflects the constraints of multilayer printing: uniform stacks require controlled ink rheology, wettability, and drying, along with reliable interlayer adhesion.\cite{tiara2022fully, Grau2016GravurePrintedElectronics, kipphan2001printmedia} Inks must remain stable during transfer and solidification, and processing temperatures must be compatible with flexible R2R substrates. Most high-performance OPV materials, including interlayers and electrodes, were developed for spin coating; only a subset meets these requirements, and even fewer can be integrated into fully printed stacks without damaging underlying layers.\cite{xue2022printing} Fully printed NFA-based OSCs have therefore not been systematically studied, and it remains unclear whether the performance mechanisms established for laboratory devices remain valid in end-to-end printed architectures.

Beyond architectural compatibility, the physical origin of performance losses upon printing is unresolved. Efficiency differences are frequently attributed to morphology,\cite{zhu2024progress} yet this explanation is rarely quantified. Printing modifies solvent evaporation, shear during ink transfer, and drying kinetics. These factors determine phase separation, molecular aggregation, and vertical composition gradients, and are further altered when chlorinated solvents are replaced by R2R-compatible alternatives.\cite{mcdowell2017organic} Thus, solvent substitution and printing do not merely change processing conditions -- they directly influence how morphology develops during film formation.\cite{zhang2025real,zhang2025revealing} At the device level, such changes can manifest as reduced photogeneration, enhanced non-radiative recombination, or limited charge transport. Without a quantitative separation of these losses, it is unclear whether the performance gap is material-limited or primarily induced by stack architecture and interfaces. For gravure-printed NFA devices, such a rigorous loss decomposition has not yet been reported.

In this work, we demonstrate state-of-the-art performance among fully R2R-compatible gravure-printed OSCs using gravure-printed PM6:Y12 active layers processed from chloroform and the non-halogenated solvent o-xylene. By combining device modelling with ellipsometry, photoluminescence, sensitive quantum efficiency, light-intensity-dependent current--voltage, and intensity-modulated photocurrent measurements, we quantitatively separate photogeneration, recombination, and charge-transport losses and link processing conditions to device physics for both chloroform and industrially compatible o-xylene systems. We show that gravure printing and non-chlorinated solvents preserve favourable bulk morphology, and that the performance gap to spin-coated references is not primarily morphological. Instead, it is governed by optical stack design and charge transport. These results establish gravure printing as a viable route for uniform thickness-controlled NFA active layers in R2R manufacturing and indicate that closing the lab-to-fab gap requires engineering of interfaces and developing printing-compatible interlayers.

\section{Results and Discussion}

\subsection{Process transfer and printing optimisation}

\begin{figure*}[tb]
    \centering
    \includegraphics[width=0.8\linewidth]{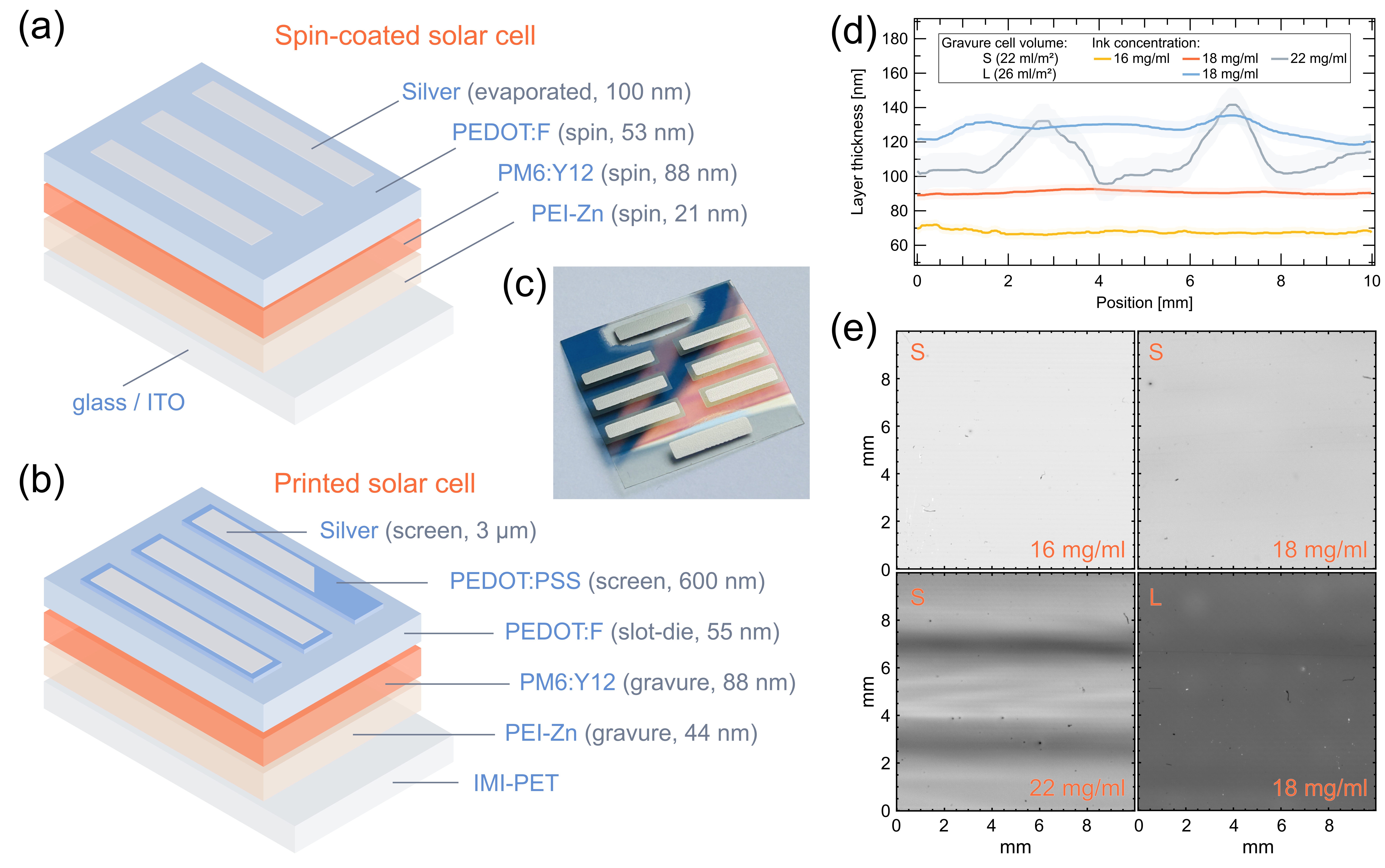}
    \caption{Layer stacks and fabrication methods for (a) spin-coated reference and (b) printed PM6:Y12 OSCs. (c) Photograph of the printed solar cell. (d) Vertical thickness profiles and (e) corresponding gray scale optical micrographs of gravure-printed o-xylene-based PM6:Y12 films using ink concentrations of 16, 18 and 22\,mg\,ml$^{-1}$ and gravure forms with 22\,ml\,m$^{-2}$ (denoted S) and 26\,ml\,m$^{-2}$ (denoted L) cell volume. Higher ink concentration leads to higher average film thickness and larger lateral thickness variations. For the 22\,mg\,ml$^{-1}$ sample pronounced non-uniformity is observed, which is overcome by using larger printing form.}
    \label{fig:lab_to_fab}
\end{figure*}

In this section, we compare the spin-coated reference architecture with the device stack compatible with R2R printing and identify process limitations that require specific optimisation for gravure printing of the active layer. As a reference, spin-coated OSCs were fabricated using PM6:Y12, the same donor--acceptor material system as employed in the printed devices. The PM6:Y12 system was selected since it is stable under air processing, and Y12 
exhibits higher solubility in non-halogenated solvents.\cite{hong2019eco} The reference spin-coated device consists of a glass/ITO substrate followed by a spin-coated electron-transport layer (ETL), the photoactive layer (AL), and a hole-transport layer (HTL), capped with an evaporated top electrode (Figure~\ref{fig:lab_to_fab}(a)). The AL had a thickness of $\sim$88\,nm. This spin-coated architecture serves as a processing reference for the material composition, solvent system, and film thickness against which the gravure-printed ALs are optimised. 

The printed device uses a R2R manufacturing-compatible stack based on PET/IMI substrate, gravure-printed ETL and AL, slot-die coated HTL, and a screen-printed electrode composed of highly conductive PEDOT:PSS and silver (Figure~\ref{fig:lab_to_fab}(b)). Gravure printing was done using a proof printing setup, where the gravure cylinder performs a single full rotation. In this configuration, ink mixing and refilling of the engraved cells with fresh ink, which is characteristic for continuous R2R printing, does not occur. Although the layer stacks and deposition methods differ reflecting laboratory versus printing-compatible architectures, the material composition and solvent system of the AL remain consistent. The photograph of the resulting small-area printed solar cell is depicted in Figure~\ref{fig:lab_to_fab}(c). 

Gravure printing imposes processing constraints that differ fundamentally from the spin coating with respect to shear conditions, wet film thickness, and drying time. During gravure printing, the ink is subjected to defined shear during cell filling, doctoring, and transfer, and it is deposited as a comparatively thick wet film that dries over extended timescales.\cite{kipphan2001printmedia} In contrast, spin-coating produces thin films under centrifugal force with rapid solvent removal. These differences lead to distinct requirements for ink viscosity, material content, and solvent evaporation rate to achieve stable ink transfer, uniform film coverage, and reproducible thickness during printing. Consequently, ink formulation and printing parameters must be specifically adjusted for gravure printing and cannot be directly transferred from spin-coating process. 

The formulation of the AL ink was systematically optimised to meet the requirements of gravure printing. Key parameters are the choice of solvent and solids concentration, which determine viscosity and drying kinetics during ink transfer and film formation.\cite{Grau2016GravurePrintedElectronics, tiara2022fully} Chloroform, widely used for laboratory-scale solar cells, provides excellent film formation under rapid solvent removal. However, its high vapour pressure and low boiling point lead to fast solvent evaporation during gravure printing and ink drying inside the gravure cells. While this behaviour can be tolerated in a single-rotation proof-printing setup, it is incompatible with continuous R2R processing, where residual ink in the cells repeatedly contacts fresh ink and must therefore remain compositionally and rheologically stable during continuous operation. For this reason, we select the less volatile o-xylene as a primary solvent. However, replacing chloroform with o-xylene changes drying dynamics and film formation, and its consequences for morphology and device performance cannot be predicted in advance. For this reason, we study both solvent systems to isolate the impact of processing.

Film thickness was tuned by adjusting the ink concentration and by varying the gravure cell volume. First, gravure printing was performed using a cylinder with a nominal cell volume of 22\,ml\,m$^{-2}$ at a web speed of 0.2\,m\,s$^{-1}$. Increasing o-xylene ink concentration from 16 to 22\,mg\,ml$^{-1}$ led to thicker ALs, as shown in Figure~\ref{fig:lab_to_fab}(d). Layer inhomogeneities were quantified from transmitted-light scans using a calibration curve to determine local film thickness (see Table~\ref{SI_tab:printed_uniformity}). Optical micrographs of the films are shown in Figure~\ref{fig:lab_to_fab}(e). At 16\,mg\,ml$^{-1}$, smooth and homogeneous $\sim$67\,nm-thick films were obtained with a coefficient of variation of $3.7\,\%$. 
Increasing the concentration to 18\,mg\,ml$^{-1}$ yielded $\sim$88\,nm-thick uniform layers with a coefficient of variation of $2.1\,\%$. At higher concentrations of 22\,mg\,ml$^{-1}$, excessive ink viscosity results in non-uniform printed layers, visible as streaked and discontinuous films with a coefficient of variation of $9.0\,\%$.

Excessive viscosity in gravure printing is known to cause poor layer quality, streaking, and mottling due to inadequate ink transfer from the engraved cells and not sufficient merging and levelling process.\cite{kipphan2001printmedia, Grau2016GravurePrintedElectronics} To overcome this, we choose ink concentration of 18\,mg\,ml$^{-1}$, but increase gravure cell volume to 26\,ml\,m$^{-2}$ to achieve thicker layers. 
The resulting films are homogeneous and 128\,nm thick. 
Although thicker AL leads to a higher short-circuit current density ($\jsc$), as shown in Figure~\ref{SI_fig:JV_thick_slim}, a lower thickness was also processed to match the spin-coated reference for a direct comparison. We therefore chose the gravure printing form with volume 22\,ml\,m$^{-2}$ and ink concentration of 18\,mg\,ml$^{-1}$ for the device studies. Under these parameters and optimised printing conditions, uniform films with an average thickness of 88\,nm were obtained over printed areas of 80\,cm$^{2}$. 

Additionally, we investigated the influence of drying conditions on device performance and identified an optimised protocol for solvent removal. A one-step box-oven drying process at 120\,°C for 5\,min  was selected as the standard condition (see Figure~\ref{SI_fig:drying})). Further fabrication details can be found in Section~\ref{SI_sec:fab} of the Supplementary Information. 

In multilayer printed devices, interfacial compatibility between sequentially deposited layers is critical to ensure uniform coverage, mechanical stability, and reproducible device performance. We therefore evaluated the interaction between the gravure-printed AL and the underlying ETL. ZnO is commonly used as ETL in laboratory-scale OSCs. However, layers printed from commercially available ZnO nanoparticle solutions exhibited poor adhesion on PET/IMI substrates and were mechanically unstable under gravure printing (see Figure~\ref{SI_fig:ETL_gravure}(a)). In contrast, PEI-Zn provided significantly improved adhesion and mechanical robustness. Adhesion tests showed that ZnO nanoparticle layers were easily damaged during handling and printing, whereas PEI-Zn layers remained intact and showed only minor surface impressions after AL deposition (see Figure~\ref{SI_fig:ETL_gravure}(b)). To further improve process safety and industrial compatibility, the PEI-Zn formulation was transferred from toxic methoxyethanol\cite{qin2020peizn_interlayer} to a non-toxic ethanol based formulation. The AL ink exhibited very good wetting on PEI-Zn, resulting in uniform coverage without de-wetting during gravure printing (see Figure~\ref{SI_fig:AL_wetting}). Blends processed from o-xylene exhibited slightly lower contact angles than those processed from chloroform, suggesting marginally improved wetting. However, both solvent systems provided adequate wettability on the PEI-Zn ETL. Together, these results demonstrate that PEI-Zn fulfills both mechanical and interfacial requirements for sequential gravure processing and is therefore a suitable ETL for the printed device stack.

To evaluate full gravure compatibility, the PEDOT:F HTL was initially deposited by gravure printing onto the printed AL. However, this approach resulted in significant mechanical damage to the underlying AL due to direct contact with the engraved gravure cylinder, as shown in Figure~\ref{SI_fig:HTL_gravure}. These impressions result in severe film failure and introduce an additional processing variable. To preserve the integrity of the gravure-printed AL and enable a direct comparison with spin-coated reference devices, the HTL was instead deposited by slot-die coating to eliminate printing-induced artefacts. Detailed optimisation of the ETL and HTL was not pursued in this work. Instead, the focus is placed on establishing a robust AL printing process within a compatible device stack. More comprehensive investigations of ETL and HTL optimisation, including alternative materials and deposition strategies, are the subject of ongoing and future studies.

The robustness of the solar cell fabrication process was evaluated in terms of reproducibility and tolerance to unavoidable process fluctuations. The resulting statistics of the solar cell parameters are summarised in Figure~\ref{SI_fig:JVstats} and Table~\ref{SI_tab:PCEstats}. For chloroform-based printed devices, an average power conversion efficiency (PCE) of 6.84\,\% was obtained, while o-xylene-based printed solar cells exhibited PCE of 6.95\,\%. The device-to-device variation further highlights the difference between the two solvent systems. The relative standard deviation of the PCE amounts to 6\,\% for chloroform-processed devices, whereas o-xylene-processed devices show a lower variation of 4\,\%. This reduced spread indicates improved process robustness for the o-xylene ink under otherwise identical printing conditions.

These results demonstrate that gravure printing of the AL does not present a fundamental barrier to R2R implementation. From a processing perspective, nothing prevents the transfer of NFA solar cells to R2R manufacturing. However, as discussed in the following sections, further development of fully printable, low-loss HTLs is required to eliminate the remaining performance gap. 
In the following, we compare the printed devices with spin-coated references from a device-physics perspective to identify the origin of the differences in their performance.

\subsection{Device physics and loss analysis}

\subsubsection{Photovoltaic performance}

To assess how gravure printing translates laboratory-scale PM6:Y12 solar cells into a R2R-compatible architecture, we systematically compare spin-coated reference devices with gravure-printed counterparts processed from chloroform and o-xylene. By keeping the material system and AL thickness comparable, differences in photovoltaic parameters can be directly attributed to processing-induced changes in optics, morphology, recombination, and charge transport.

\begin{figure}[b]
    \centering
    \includegraphics[width=0.8\linewidth]{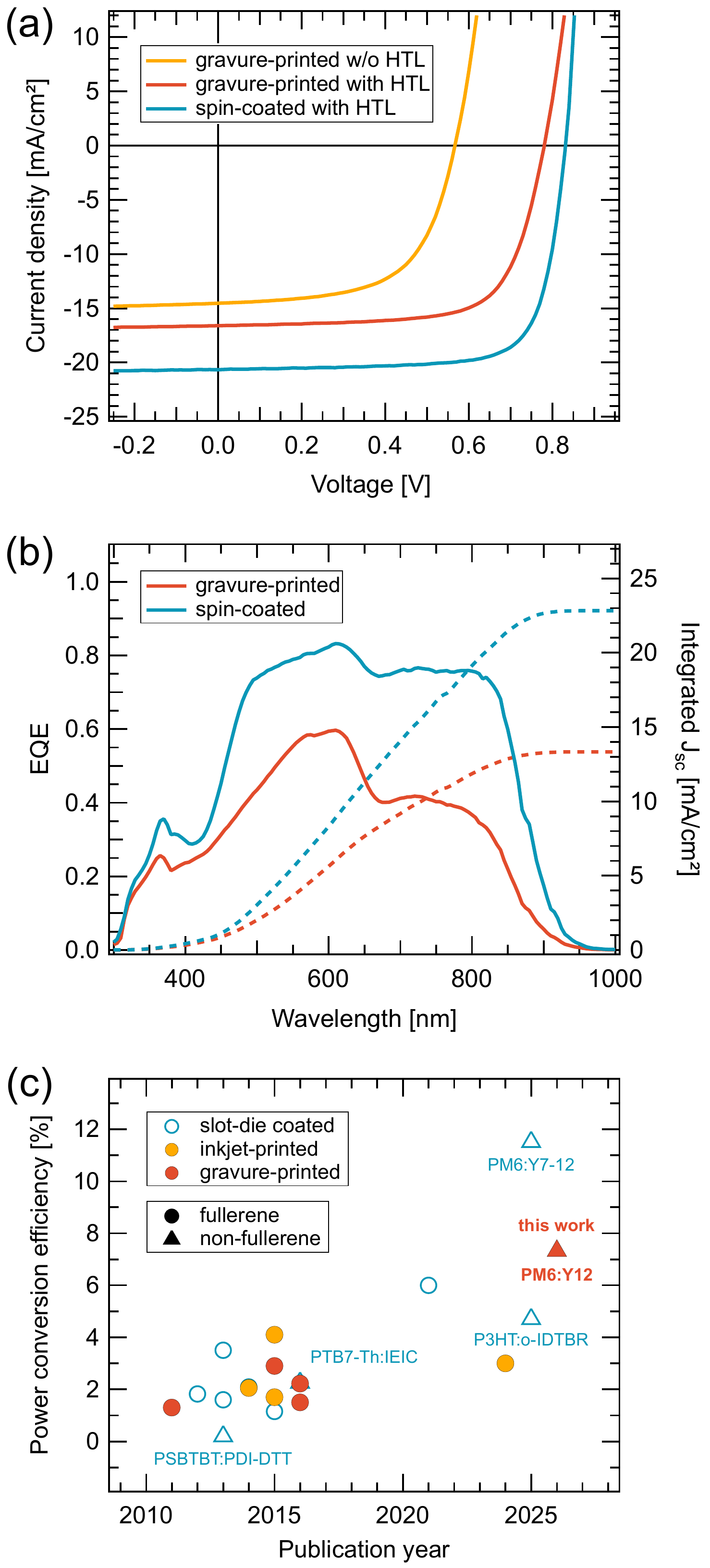}
    \caption{
    (a) JV characteristics, and 
    (b) EQE spectra of gravure-printed and spin-coated o-xylene-based PM6:Y12 devices. JV curves were measured under 1\,sun equivalent illumination provided by Wavelabs~LS2 solar simulator (for spectrum, see Figure~\ref{SI_fig:spectrum}). 
    (c) PCE values of R2R-compatible OSCs from literature compared to current work, highlighting different AL deposition methods (slot-die coating, inkjet printing, and gravure printing) and active-layer systems for NFA OSCs. The parameters and references are listed in Table~\ref{SI_tab:PCE_lit}.}
    \label{fig:JV_EQE}
\end{figure}

We first compare the spin-coated reference with gravure-printed devices processed from o-xylene. The current-density--voltage (JV) characteristics and external quantum efficiency (EQE) spectra are presented in Figures~\ref{fig:JV_EQE}(a) and (b), respectively, and solar cell parameters are summarised in Table~\ref{tab:PCE}. The fully printed architecture exhibits the largest performance loss relative to the spin-coated baseline. PEDOT:F formulation cannot be screen printed with sufficient film quality, and gravure deposition damages the underlying AL. Consequently, no fully printable HTL is currently available that can be deposited directly on the AL without compromising device performance.

To enable proper energy-level alignment between PM6 and the top electrode and increase the open-circuit volatge ($\Voc$), we introduce a slot-die coated PEDOT:F interlayer. This intermediate architecture improves device performance compared to the fully printed stack, yielding a PCE of 7.28\,\% for devices processed from o-xylene and 7.32\,\% for those processed from chloroform. Nevertheless, $\jsc$ and $\Voc$ remain below the spin-coated reference, indicating that the printed stack still introduces additional losses. 

\begin{table}[t]
\caption{Representative solar cell parameters for gravure-printed and spin-coated devices based on chloroform and o-xylene. The values of $\jsc^\mathrm{*}$ and PCE$^\mathrm{*}$ have been corrected for the AM1.5G spectrum and PEDOT:PSS area.}
\centering
\small
\renewcommand{\arraystretch}{1.15}
\begin{tabular}{lccccc}
\hline
Device type & $\Voc$ & $\jsc$ & $\jsc^\mathrm{*}$ & $FF$ & PCE$^\mathrm{*}$ \\
& (mV) & (mA\,cm$^{-2}$) & (mA\,cm$^{-2}$) & (\%)  & (\%) \\
\hline
\multicolumn{6}{c}{PM6:Y12 from o-xylene} \\
\hline
printed w/o HTL & 567 & 14.54 & 11.66 & 60.61 & 4.01 \\
printed with HTL & 781 & 16.60 & 13.32 & 69.99 & 7.28 \\
spin-coated  & 831 & 20.67 & 22.83 & 75.81 & 14.38 \\
\hline
\multicolumn{6}{c}{PM6:Y12 from chloroform} \\
\hline
printed with HTL  & 801 & 16.66 & 13.36 & 68.42 & 7.32 \\
spin-coated  & 834 & 20.36 & 22.54 & 72.52 & 13.63 \\
\hline
\multicolumn{6}{l}{\footnotesize $^\ast$ AM1.5G and area-corrected $\jsc$ and PCE values.}
\end{tabular}\label{tab:PCE}
\end{table}

The photovoltaic parameters in Table~\ref{tab:PCE} include $\jsc$ measured under the LS2 solar simulator and corrected values. Since the LS2 spectrum deviates from AM1.5G (Figure~\ref{SI_fig:spectrum}), $\jsc$ was recalculated by integrating the measured EQE with the AM1.5G photon flux. For printed devices, integrated EQE measured through a defined aperture also eliminates overestimation of $\jsc$ arising from the PEDOT:PSS layer extending beyond the silver electrode (Figure~\ref{fig:lab_to_fab}(b)), which otherwise allows additional lateral charge collection. This procedure yields the current density corresponding to the geometrically defined active area.

We next compare devices processed from chloroform and o-xylene. The JV curves and EQE of chloroform-based solar cells are presented in Figure~\ref{SI_fig:CF_JV_EQE}. As for o-xylene, the photovoltaic performance of gravure-printed devices is lagging behind the spin-coated reference. Photovoltaic parameters in Table~\ref{tab:PCE} reveal that in both fabrication routes, chloroform leads to higher $\Voc$, while o-xylene yields improved fill factor ($\FF$). In gravure-printed devices, these trends are reproducible across multiple devices, as shown in the statistical analysis (Figure~\ref{SI_fig:JVstats} and Table~\ref{SI_tab:PCEstats}). 

Overall, the comparison reveals two systematic effects: (i) printing reduces $\jsc$, $\Voc$, and $\FF$ compared to spin coating, and (ii) solvent choice primarily influences $\Voc$ and $\FF$ for both fabrication routes. In the following sections, we analyse the physical origin of these differences by separating optical and morphology-induced losses affecting $\jsc$, recombination-related voltage losses governing $\Voc$, and transport limitations determining $\FF$.

To place our results in the context of scalable organic photovoltaics, Figure~\ref{fig:JV_EQE}(c) compares the PCE values of reported R2R-compatible OSCs. The comparison includes only devices in which all layers were deposited by scalable techniques; studies relying on laboratory-scale processes such as spin-coating, blade coating, or thermal evaporation were excluded. Within this set, previous reports include devices with slot-die coated\cite{yu2012silver, liu2013all, helgesen2013slot, hoesel2013fast, cheng2014comparison, galagan2015roll, liu2016roll-coating, miranda2021efficient,feroze2025long,jayaraman2025flexible} and printed\cite{huebler2011printed,jung2014all-inkjet,eggenhuisen2015high, vilkman2015gravure, kapnopoulos2016fully,vak2016reverse, Steinberger2024} active layers based on both fullerene and non-fullerene systems. In this context, our PM6:Y12 device represents the \emph{first fully R2R-compatible gravure-printed OSC employing an NFA active layer}, and it achieves \emph{the highest efficiency} among reported gravure-printed OSCs as well as among fully R2R-compatible printed OSCs overall.

\subsubsection{Optical photogeneration losses}

Despite the competitive performance of the gravure-printed PM6:Y12 devices among R2R-compatible OSCs, the printed stack still exhibits noticeable losses relative to the spin-coated reference. Since the reduction in $\jsc$ represents the largest performance difference between gravure-printed and spin-coated devices, we first analyse purely optical contributions to photogeneration. We separate intrinsic absorption-related effects of the AL from stack-induced optical losses arising from the modified device architecture.
\medskip

\paragraph{Absorption strength and optical anisotropy.}

\begin{figure*}[t]
    \centering
    \includegraphics[width=0.8\linewidth]{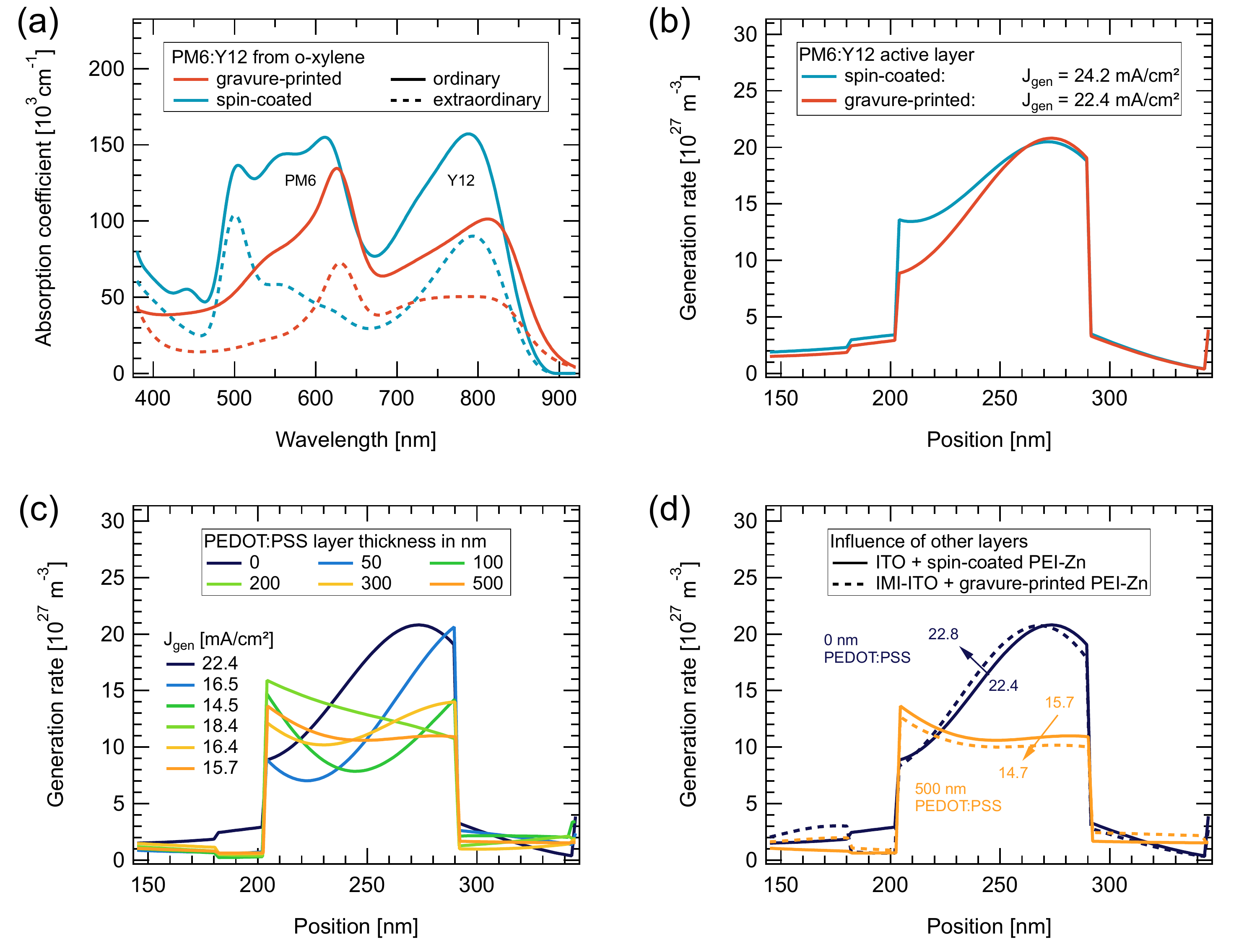}
    \caption{(a) Ordinary and extraordinary absorption coefficients for spin-coated and gravure-printed PM6:Y12 films determined by VASE, revealing optical anisotropy and showing reduced absorption strength in printed films. (b-d) Optical transfer-matrix modelling of the generation current density $\jgen$: (b) influence of the AL optical constants on photogeneration; (c) effect of PEDOT:PSS thickness on $\jgen$; (d) comparison of different bottom electrode and ETL configurations.}
    \label{fig:transfermx}
\end{figure*}

We measured spectrally-resolved absorption coefficients of gravure-printed and spin-coated PM6:Y12 films processed from o-xylene using variable-angle spectroscopic ellipsometry (VASE). The optical response was separated into the ordinary ($\alpha_{xy}$) and extraordinary($\alpha_{z}$) components, corresponding to in-plane and out-of-plane polarisation, respectively (Figure~\ref{fig:transfermx}(a)). Both gravure-printed and spin-coated films exhibit pronounced optical anisotropy, with systematically higher in-plane than out-of-plane absorption across the main absorption bands of PM6 and Y12. We observe an overall reduced absorption strength in the printed films compared to the spin-coated reference, while preserving the characteristic spectral shape of the PM6:Y12 blend. 

We quantify the optical anisotropy by extracting the orientation order parameter $S$ and the average molecular orientation angle $\phi$ from the extinction coefficients $k$. The orientation order parameter is defined as
\begin{equation}
S = \frac{1}{2}\left(3\sin^2\phi - 1\right) = \frac{k_{z} - k_{xy}}{k_{z} + 2k_{xy}},
\end{equation}
where $S = 0$ ($\phi = 35.4^\circ$) corresponds to random molecular orientation, $S = -0.5$ indicates molecules oriented parallel to the substrate surface, and $S = 1$ corresponds to molecules oriented perpendicular to the surface.\cite{gordan2004anisotropy} The analysis assumes that the dominant optical transition dipole moments are oriented parallel to the polymer backbone for PM6 and parallel to the molecular plane for Y12, consistent with established models and literature reports.\cite{kroh2022identifying,eller2024tackling}

For Y12, the analysis was performed over the spectral range above 670\,nm, while for PM6 the range from 460 to 670\,nm was used. The extracted values show that PM6 exhibits a slightly stronger face-on orientation in the printed films compared to the spin-coated reference ($S = -0.24 \pm 0.01$, $\phi = 24.4 \pm 0.5^\circ$ vs.\ $S = -0.22 \pm 0.01$, $\phi = 25.5 \pm 0.5^\circ$), whereas Y12 shows a marginally stronger face-on orientation in the spin-coated films ($S = -0.20 \pm 0.01$, $\phi = 26.8 \pm 0.5^\circ$ vs.\ $S = -0.17 \pm 0.01$, $\phi = 28.0 \pm 0.5^\circ$). These differences are small but indicate subtle changes in molecular orientation induced by the deposition method.

From an application perspective, the observed optical anisotropy is not expected to limit device operation, as solar cells are exposed to a broad distribution of illumination angles under both outdoor and indoor conditions.
\medskip

\paragraph{Layer-stack-induced photogeneration losses.}

To quantify optical contributions to the reduced $\jsc$ originating from the layer stack, we performed optical simulations using experimentally determined refractive indices and extinction coefficients. The simulations were carried out using a transfer-matrix method implemented in the OghmaNano device simulation software (see Section~\ref{SI_sec:optical_sim} for details).\cite{mackenzie2012extracting,dattani2014general,liu2015organic,bickerdike2025unravelling}
From the simulated position-dependent generation rate $G(x)$, the generation current density $\jgen$ was obtained by integrating $G(x)$ over the AL thickness. The results are summarised in Figures~\ref{fig:transfermx}(b)--(d).
 
First, we isolate the impact of the AL optical constants, by keeping the device stack identical to the spin-coated reference and varying only the complex refractive index of the AL (Figure~\ref{fig:transfermx}(b)). Replacing the spin-coated optical constants with those of the printed AL reduces $\jgen$ from 24.2 to 22.4\,mA\,cm$^{-2}$, consistent with the reduced absorption strength of the printed AL discussed above.

We then access the influence of a PEDOT:PSS overlayer with thicknesses between 0 and 500\,nm while retaining the printed AL optical constants (Figure~\ref{fig:transfermx}(c)). Increasing the PEDOT:PSS thickness leads to a pronounced and non-monotonic decrease in $\jgen$, reaching values as low as $\sim$16\,mA\,cm$^{-2}$ for thick layers. This reduction is attributed to optical interference effects and parasitic absorption within the PEDOT:PSS layer, demonstrating that thick printed transport layers can introduce substantial optical generation losses even when the AL itself is unchanged. In real devices, an additional loss is likely originating from rough PEDOT:PSS/silver interface. 

Finally, we evaluate the effect of the bottom electrode and ETL on photogeneration by comparing ITO with spin-coated PEI-Zn to IMI/ITO with printed PEI-Zn (Figure~\ref{fig:transfermx}(d)). In both cases, for thin and thick PEDOT:PSS layers, we observe only minor differences in the generation profile and in $\jgen$, with variations not exceeding 1\,mA\,cm$^{-2}$. Overall, our transfer-matrix simulations demonstrate that optical losses associated with the printed AL and, in particular, with thick printed PEDOT:PSS layers contribute significantly to the reduced $\jsc$ of printed devices. These losses are purely optical in origin and must be considered alongside electrical loss mechanisms when translating laboratory device stacks to printing-compatible architectures.

\subsubsection{Active-layer morphology and vertical composition}

We next examine bulk morphology on the exciton length scale by analysing the exciton quenching efficiency (see Section~\ref{SI_sec:PL} for details). Under excitation at 500\,nm, the photoluminescence of the blends originates predominantly from Y12 excitons. We quantify exciton quenching by integrating the photoluminescence spectra over the wavelength range from 830 to 1200\,nm and normalising the integrated intensity by the optical absorptance at the excitation wavelength to account for differences in film thickness and absorption strength. 

All samples exhibit very high exciton quenching efficiencies. 
Spin-coated films processed from chloroform and o-xylene show quenching ratios of 96.3\,\% and 96.9\,\%, respectively, while the corresponding printed films yield slightly lower values of 94.9\,\% (chloroform) and 95.0\,\% (o-xylene). These results indicate that exciton harvesting is efficient in both fabrication routes, with only a small residual loss that is more pronounced in printed devices. 

The reduced quenching observed for printed samples compared to spin-coated references suggests slightly less effective exciton harvesting, consistent with a larger donor--acceptor domain size in the printed films. We attribute this trend to the slower drying kinetics during gravure printing, which allow increased phase separation during film formation, whereas the rapid solvent removal during spin coating potentially suppresses domain growth.\cite{Jiang2022} Overall, exciton quenching does not represent a major limitation for device performance, although small morphology-related differences may contribute to the reduced EQE and $\jsc$ in printed devices.

A solvent-dependent trend is also observed. Despite its slower evaporation, the films processed from o-xylene exhibit slightly higher quenching efficiencies compared to chloroform, indicating that thermodynamic factors such as improved donor--acceptor miscibility and solubility in o-xylene outweigh kinetic effects and promote finer effective mixing.

To further assess whether bulk aggregation contributes to the observed performance differences, we compare normalised absorbance spectra of PM6 and Y12 for printed and spin-coated films (Figure~\ref{SI_fig:abs}). Both materials show only minor solvent- and printing-induced variations in their absorbance spectra. When comparing printed films, o-xylene-processed samples show a small redshift for both components relative to chloroform-processed films. For PM6, this redshift is accompanied by a slightly increased 0--0 to 0--1 vibronic peak ratio, indicating marginally enhanced chain ordering or stronger interchain coupling. For Y12, the o-xylene-processed printed film shows a weak redshift of the absorption maximum together with a slightly enhanced low-energy tail, consistent with modestly stronger aggregation or increased energetic disorder. 

Taken together, exciton quenching and normalised absorption data indicate that differences in bulk morphology are subtle and unlikely to dominate device performance. Contact-angle analysis of PM6:Y12 films (Section~\ref{SI_sec:surfaceE}) further reveals preferential PM6 enrichment at the air interface and Y12 accumulation toward the PEI-Zn ETL layer with high surface energy, forming an interface-selective vertical composition that promotes charge extraction. These results show that the PM6:Y12 system intrinsically forms a favourable bulk and vertical morphology. The performance gap between spin-coated and printed devices therefore cannot be attributed primarily to severe morphological degradation of the donor--acceptor network.

\subsubsection{Radiative and non-radiative voltage losses}

\begin{figure*}[bt]
    \centering
    \includegraphics[width=0.8\linewidth]{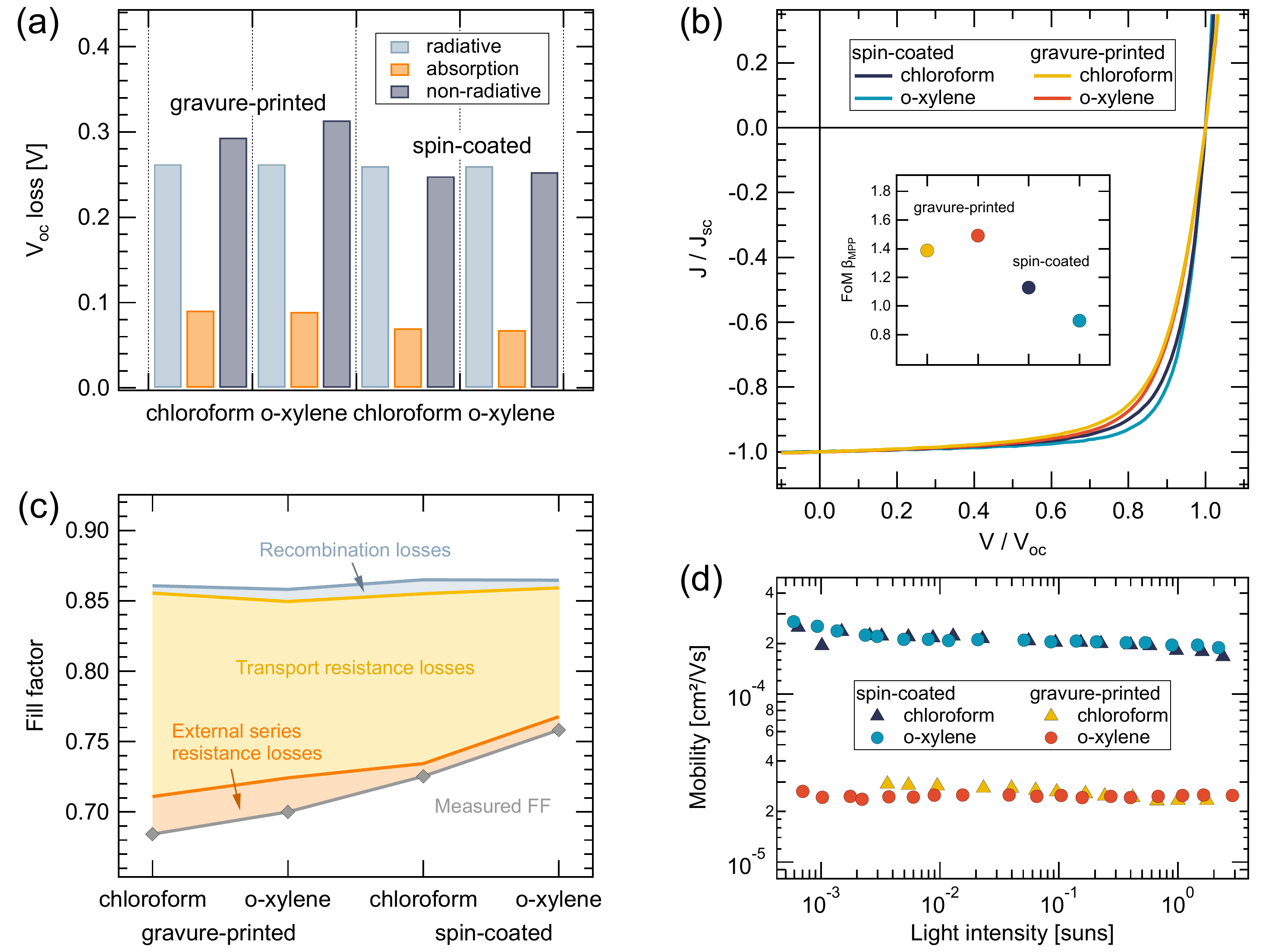}
    \caption{Various loss parameters in spin-coated and gravure-printed PM6:Y12 solar cells processed from chloroform and o-xylene. (a) Decomposition of $\Voc$ losses into the Shockley--Queisser radiative limit, the additional radiative loss associated with non-ideal absorption, and the non-radiative recombination loss. (b) Analysis of $\FF$ losses. JV characteristics are normalised to highlight differences in $\FF$; the inset shows the extracted transport figure of merit $\beta_\mathrm{MPP}$. (c) $\FF$ losses separated into contributions related to recombination (shaded grey), transport (shaded yellow), and external series resistance (shaded orange). The solid lines indicate the fill factors expected in the absence of the corresponding losses. (d) Charge-carrier mobilities derived from IMPS measurements. }
    \label{fig:voc_ff_losses}
\end{figure*}

To identify the origin of the reduced open-circuit voltage in printed devices, we analyse the $\Voc$ losses using a detailed-balance framework based on the photovoltaic external quantum efficiency.\cite{rau2007reciprocity,yao2015quantifying} The analysis procedure is described in Section~\ref{SI_sec:Voc} of the Supplemental Material, and the resulting loss contributions are summarised in Figure~\ref{fig:voc_ff_losses}(a) (for values see Table~\ref{SI_table:voc_losses}). The total voltage loss, $\Eg/e - \Voc$, is decomposed into three contributions: the unavoidable radiative loss in the Shockley--Queisser limit, an additional radiative loss associated with non-ideal absorption, and the non-radiative recombination loss.

As shown in Figure~\ref{fig:voc_ff_losses}(a), the unavoidable radiative loss is comparable for all devices, reflecting similar optical gaps extracted from $\EQE$. Printed devices exhibit a larger voltage loss related to absorption compared to spin-coated references. This increase is primarily caused by the reduced overall photogeneration rate in the printed architecture. In particular, the thick PEDOT:PSS layers required for fully printed electrodes strongly attenuate the optical field reflected from the silver electrode. As a result, both the effective absorption and the EQE are reduced, directly lowering the radiative open-circuit voltage. A secondary contribution arises from a slightly less steep absorption edge for printed devices -- also caused by the thick PEDOT:PSS layer -- leading to a higher calculated energy gap. This effect is smaller than the loss associated with reduced optical generation. However, it is important to note that the optical gap determined from the absorption edge is not only a function of material, and will be affected by interference within the layer stack.

In addition to the increased radiative voltage loss, printed devices show a larger non-radiative voltage loss than spin-coated references (Figure~\ref{fig:voc_ff_losses}(a)), indicating an increased contribution of defect-assisted or interfacial recombination pathways in the printed devices. We attribute this effect to the modified device interfaces required for fully printed architectures. In particular, printed transport layers and sequential printing steps lead to rougher and more disordered interfaces at the AL,\cite{Grau2016GravurePrintedElectronics,brumm2021ink, wang2024manipulating} introducing additional non-radiative recombination channels. 

A solvent-dependent difference in $\Voc$ is observed for both spin-coated and printed devices, with chloroform-processed samples consistently exhibiting a higher $\Voc$ than those processed from o-xylene (Tables~\ref{tab:PCE} and \ref{SI_tab:PCEstats}). Because this trend is present for both fabrication routes, it reflects solvent-induced differences in the AL rather than effects of the printed device stack. The lower $\Voc$ in o-xylene-processed devices originates from increased non-radiative voltage losses, indicating additional non-radiative recombination channels compared to chloroform. Since the recombination ideality factor $\nid$ is comparable for both solvents (Figure~\ref{SI_fig:I4-nid}), this increase cannot be attributed to enhanced energetic disorder but instead points to a higher density of recombination-active interfaces in the o-xylene-processed films, consistent with slightly higher exciton quenching ratios discussed previously. 

To better understand the experimental trends, we combined device modelling with machine learning. We generated a large synthetic dataset of devices using the same optical stack as the experimental structures, including the spin-coated reference architecture, while randomly varying the electrical parameters. This procedure produced a broad range of physically consistent JV curves paired with known microscopic parameters. A machine-learning model was then trained to learn the inverse mapping from illuminated JV characteristics to the underlying device parameters. Specifically, we used a residual feedforward neural network architecture,\cite{Rumelhart1986} where shortcut connections between layers\cite{he2015deepresiduallearningimage} improve training stability and enable robust extraction of correlated parameters. The trained network directly infers charge transport and recombination properties from steady-state JV curves, providing a data-driven parameter inversion consistent with the calibrated device model.

\begin{table}[t]
\caption{Charge generation and recombination parameters for printed and spin-coated PM6:Y12 devices processed from o-xylene and chloroform, extracted from simulations: the generation rate $G$ from optical simulations and the recombination lifetime $\tau$ from machine learning. The effective second-order recombination coefficient $k_2$ and the effective density of states $N_0$ were calculated as described in the text.}
\centering
\small
\renewcommand{\arraystretch}{1.15}
\begin{tabular}{lcccc}
\hline
Device type & \quad$\bar{G}$   & \quad$\tau$ & \quad$k_2$ & \quad$N_0$\\
& \quad(s$^{-1}$\,cm$^{-3}$) & \quad(\textmu{}s) & \quad(cm$^{3}$\,s$^{-1}$) & \quad(cm$^{-3}$)\\
\hline
\multicolumn{5}{c}{PM6:Y12 from o-xylene} \\
\hline
printed     & \quad$9.95\cdot 10^{21}$ & \quad$3.25$  & \quad$9.51\cdot 10^{-12}$ & \quad$1.16\cdot10^{22}$ \\
spin-coated & \quad$1.73\cdot 10^{22}$ & \quad$3.09$  & \quad$6.06\cdot 10^{-12}$ & \quad$3.81\cdot10^{21}$ \\
\hline
\multicolumn{5}{c}{PM6:Y12 from chloroform} \\
\hline
printed     & \quad$9.95\cdot 10^{21}$ & \quad$2.64$  & \quad$1.44\cdot 10^{-11}$ & \quad$6.53\cdot 10^{21}$ \\
spin-coated & \quad$1.73\cdot 10^{22}$ & \quad$2.28$  & \quad$1.11\cdot 10^{-11}$ & \quad$2.65\cdot 10^{21}$ \\
\hline
\end{tabular}\label{tab:rec_par}
\end{table}

Recombination lifetimes $\tau$ extracted from the machine-learning analysis were used to estimate the effective second-order recombination coefficient $k_2$ (Table~\ref{tab:rec_par}), calculated as $k_2 = 1/(\bar{G}\tau^2)$. The spatially averaged generation rate $\bar{G}$ was obtained from the optical simulations discussed above and assumed to be identical for o-xylene and chloroform. Printed devices exhibit systematically higher $k_2$ values than the spin-coated references, indicating stronger nongeminate recombination, consistent with the larger $\Voc$ losses observed experimentally.

Under open-circuit conditions, the charge-carrier density $n$ is determined by the balance between generation and recombination. The quasi-Fermi level splitting depends on $n$ approximately as $\Voc \approx (2\kT/e)\ln(n/n_0)$, where the charge-carrier densities under illumination and in the dark follow $n = \bar{G}\tau$ and $n_0 = N_0\exp\left(-\Eg/(2\kT)\right)$, respectively. From these relations, we can estimate the effective density of states $N_0$. In organic semiconductors, the electronic states are largely localised and therefore act predominantly as trap states.\cite{bassler1993charge,saladina2023power} Consistent with the experimental $\Voc$ trend, the extracted values in Table~\ref{tab:rec_par} suggest a higher trap density for o-xylene compared to chloroform and for printed devices compared to spin-coated ones. While the absolute values should be interpreted with caution due to the small differences, the observed trends are robust. Finally, a significant fraction of the nonradiative losses may originate from interface recombination, which would reduce the apparent contribution from bulk recombination. This contribution cannot be resolved with the present analysis. 

These observations provide insight into the experimentally observed $\Voc$ losses and highlight design aspects of the printed device architecture that can be further optimised. The thick PEDOT:PSS interlayer in the printed architecture prevents direct contact between the AL and the screen-printed silver electrode. However, it introduces an optical penalty by attenuating back-reflection from the silver electrode, thereby reducing photogeneration and increasing the radiative $\Voc$ loss. A direct route to improve $\Voc$ is therefore to reduce the optical thickness and parasitic absorption of this protective layer while maintaining its barrier function, for example by using thinner highly conductive PEDOT:PSS formulations, alternative transparent barrier/HTL materials, or modified printed electrode stacks that enable direct deposition without damaging the AL.

In parallel, the larger non-radiative voltage losses in printed devices motivate further interface optimisation, particularly reducing roughness and defect density at the AL interfaces through improved wetting control, interlayer smoothing, or lower-stress electrode printing. Finally, the solvent dependence indicates that o-xylene processing introduces additional non-radiative recombination pathways, suggesting that further optimisation of film formation in o-xylene will be required to suppress recombination-active interfaces while retaining the manufacturing window required for gravure printing.

\subsubsection{Charge transport and fill factor losses}

To analyse the origin of the reduced $\FF$ in printed devices, we evaluate the JV characteristics using the transport-resistance framework.\cite{wurfel2015impact,neher2016new,saladina2025transport} The detailed methodology is described in Section~\ref{SI_sec:I4-JV}. Pseudo-JV curves were constructed to represent the device response in the absence of transport resistance, using the recombination ideality factor $\nid$ and a generation current density $\jgen$ estimated at $-0.5$\,V. Comparison of these pseudo-JV curves with the experimentally measured characteristics in Figure~\ref{SI_fig:I4-JV} highlights the voltage losses introduced by charge transport.

To compare $\FF$ losses independently of variations in $\jsc$ and $\Voc$, normalised JV curves are shown in Figure~\ref{fig:voc_ff_losses}(b), where current density and voltage are scaled by $\jsc$ and $\Voc$, respectively. In this representation, printed devices show a reduced curvature compared to the spin-coated reference, indicating increased transport-related losses. The magnitude of transport resistance at the maximum power point is quantified using the figure of merit $\beta_\mathrm{MPP}$, which was obtained by the iterative procedure using Eq.~\eqref{SI_eq:betampp}.\cite{saladina2025transport} The extracted $\beta_\mathrm{MPP}$ values, summarised in the inset of Figure~\ref{fig:voc_ff_losses}(b), are consistently larger for printed devices, reflecting slower charge extraction under operating conditions. Additional insight into charge extraction is obtained from biased EQE measurements in Figure~\ref{SI_fig:biasedEQE}. The EQE spectra retain spectral shape with applied bias, indicating that photogeneration is largely field independent. Printed devices exhibit a stronger overall field dependence than spin-coated references, especially the chloroform-processed one, consistent with the conclusion that reduced charge transport is the primary origin of the lower $\FF$. Earlier, we demonstrated that PM6:Y12 forms a favourable vertical gradient that promotes selective charge extraction. Therefore, the reduced $\FF$ in printed devices is unlikely to originate from interfacial charge selectivity and is instead governed by charge transport in the bulk of the active layer.

In general, the $\FF$ is governed by recombination and resistive effects arising from the external series resistance $\Rext$ and the transport resistance $\Rtr$, which originate from voltage drops at the contacts and charge transport through the active layer, respectively.\cite{schiefer2014determination,neher2016new} Figure~\ref{fig:voc_ff_losses}(c) decomposes the resulting $\FF$ losses into recombination-, transport-, and external series resistance-related contributions. The orange line represents the $\FF$ expected in the absence of $\Rext$, the yellow line additionally excludes $\Rtr$, while the blue line indicates the maximum achievable $\FF$ at the measured $\Voc$. The external-series-resistance-free $\FF$ was obtained by correcting the applied voltage for the voltage drop across $R_\mathrm{ext}$, determined from the slope $\der V/\der J$ of the forward-bias region (Figure~\ref{SI_fig:I4-JV}(b)). The remaining reference values were calculated using the Green equation,\cite{green_accuracy_1982}
\begin{equation}\label{eq:FF_green}
\begin{aligned}
\FF &= \frac{\uoc - \ln\!\left(\uoc + 0.72\right)}{\uoc + 1} , \\
\uoc &= \frac{e\Voc}{\nid\kT} .
\end{aligned}
\end{equation}
For the maximum $\FF$, $\nid$ was set to unity, corresponding to the ideal recombination limit. 

The analysis shows that recombination contributes least to the $\FF$ losses, with similar values for printed and spin-coated devices. The next-largest contribution arises from external series resistance, which accounts for a substantial fraction of the $\FF$ reduction in printed solar cells. To identify its origin, we fabricated devices in which only the ETL and AL were gravure-printed, while the HTL and electrode were identical to those in the spin-coated reference. The resulting $\Rext$ values (Figure~\ref{SI_fig:I4-JV}(b)) indicate that approximately 35\,\% of the resistance increase originates from the HTL side and 65\,\% from the ETL side. On the ETL side, handling of the flexible IMI-PET substrate may induce microcracks in the electrode. High gravure pressure during ETL printing can further damage the IMI-PET, and gravure imprinting during AL printing can degrade the ETL/AL interface. On the HTL side, the printed PEDOT:PSS electrode has lower conductivity than metal contacts and forms rough interfaces with the PEDOT:F interlayer and silver electrode; in addition, the acidic PSS component may interact with silver. Together, these effects increase the overall series resistance and contribute substantially to the $\FF$ losses in gravure-printed devices.

After removing the $\Rext$ contribution, the remaining gap between the orange and yellow curves in Figure~\ref{fig:voc_ff_losses}(c) reflects transport losses in the photoactive layer. These account for the largest share of the $\FF$ loss in printed solar cells and contribute substantially to the performance difference between the two fabrication routes. Transport losses are slightly smaller for o-xylene than for chloroform, suggesting that ALs processed from o-xylene provide more favourable transport pathways. 

To quantify charge transport, we use intensity-modulated photocurrent spectroscopy (Section~\ref{SI_sec:IMPS}). Figure~\ref{fig:voc_ff_losses}(d) shows that the effective charge-carrier mobility in printed devices is nearly one order of magnitude lower than in spin-coated references across more than three decades in light intensity. The mobilities extracted from machine learning are consistent in magnitude with the experimentally derived values (Table~\ref{tab:ml_core_parameters}). 
For identical recombination rates, such a mobility reduction would result in a substantially lower $\FF$. However, printed devices also exhibit a strongly reduced photocurrent, mainly due to optical losses from the thick PEDOT:PSS interlayer. The resulting lower photogeneration reduces the steady-state charge-carrier density and recombination rate, consistent with the reduced $\Voc$ observed in printed devices. Consequently, the experimentally observed $\FF$ reduction is smaller than expected from mobility alone.

To isolate the intrinsic charge-carrier mobility limitation to $\FF$ in the printed device, we assume the same photogeneration and charge-carrier density as in the spin-coated reference. The $\FF$ is then predicted using the Green equation, Eq.~\eqref{eq:FF_green}, where the recombination ideality factor is replaced by the apparent factor $\nid + \beta_{\mathrm{MPP}}$, accounting for recombination and transport resistance but excluding the external series resistance.\cite{saladina2025transport} This yields a predicted $\FF$ of 52.75\,\% for the printed device (see Section~\ref{SI_sec:I4-JV}). As a consistency check, applying the same procedure to the spin-coated reference gives a predicted $\FF$ of 78.01\,\%, in close agreement with the external-resistance-free value of 76.76\,\%. These results indicate that, once the optical stack is optimised and photogeneration increases, charge transport will become the dominant performance limitation in printed devices.

\section{Conclusion}
 
In this work, we establish a practical pathway for transferring OSCs from spin coating to gravure printing, achieving record performance for gravure-printed OSCs within fully R2R-compatible device stacks. We directly compare gravure-printed PM6:Y12 devices based on chloroform and o-xylene to their spin-coated counterparts to isolate the influence of solvent and deposition method. We demonstrate that careful optimisation of ink formulation and interlayer compatibility enables uniform printed active layers with reproducible performance, while highlighting the current bottleneck imposed by the absence of a fully printable, low-loss hole-transport layer. 

The subsequent device-physics analysis identifies the dominant origins of the performance gap. The largest loss arises from reduced $\jsc$, primarily caused by stack-induced optical losses associated with thick screen-printed PEDOT:PSS interlayer. Bulk morphology differences are comparatively small and do not fundamentally limit performance. The reduced $\Voc$ in printed devices results from both increased radiative losses linked to lower photogeneration and enhanced non-radiative recombination. Finally, the lower $\FF$ is mainly governed by transport resistance rather than recombination. Solvent choice further modulates non-radiative losses and charge transport, with chloroform yielding higher $\Voc$ and o-xylene slightly improving charge extraction. Overall, our results disentangle processing-induced optical, morphological, recombination, and transport losses, providing a clear roadmap for closing the performance gap between laboratory-scale and fully printed organic solar cells.

\section*{Acknowledgements}
We thank the Deutsche Forschungsgemeinschaft (DFG) for funding this work (Research Unit FOR~5387 POPULAR, project no.~461909888).

\bibliographystyle{apsrev4-2}
\bibliography{references}

\clearpage
\onecolumngrid 
\graphicspath{{figures/}}
\renewcommand{\thepage}{S\arabic{page}}  
\renewcommand{\thesection}{S\arabic{section}}
\renewcommand{\thetable}{S\arabic{table}}   
\renewcommand{\thefigure}{S\arabic{figure}}
\renewcommand{\theequation}{S\arabic{equation}}
\renewcommand{\figurename}{Figure}
\renewcommand{\tablename}{Table}
\setlength{\parskip}{0.3cm}
\setlength{\parindent}{0pt}
\setcounter{page}{1}
\setcounter{figure}{0}
\setcounter{table}{0}
\setcounter{equation}{0}
\setcounter{section}{0}

\begin{center}
    \large
    Supplementary Information
    \vspace{0.2cm}
\end{center}

\maketitle

\section{Materials and device fabrication}\label{SI_sec:fab}

\subsection{Materials} 

Conductive transparent electrode ITO-Ag-ITO (IMI) (sheet resistance $8.0\,\Omega/\square \pm 0.5\,\Omega/\square$) on polyethylene terephthalate foil was obtained from ASKA GmbH. 
Branched polyethylenimine (PEI) (Mw\,$\sim$25.000), zinc acetate dihydrate, ethanol, $\geq 99.9\,\%$, chloroform $\geq 99.5\,\%$ and o-xylene $\geq 98.0\,\%$ were obtained from Sigma-Aldrich. PEDOT:PSS EL-P5015 ink was obtained from Orgacon (sheet resistance $190\,\Omega/\square$). BM-HTL-1, Poly[(2,6-(4,8-bis(5-(2-ethylhexyl)-4-fluorothiophen-2-yl)-benzo[1,2-b:4,5-b’]dithiophene))-alt-(5,5-(1’,3’-di-2-thienyl-5’,7’-bis(2-ethylhexyl)benzo[1’,2’-c:4’,5’-c’]dithiophene-4,8-dione))] (PM6) and 2,2'-((2Z,2'Z)-((12,13-bis(2-butyloctyl)-3,9-diundecyl-12,13-dihydro-[1,2,5]thiadiazolo[3,4-e]thieno[2",3":4',5']thieno[2',3':4,5]pyrrolo[3,2-g]thieno[2',3':4,5]thieno[3,2-b]indole-2,10-diyl)bis(methanylylidene))bis(5,6-difluoro-3-oxo-2,3-dihydro-1H-indene-2,1-diylidene))dimalononitrile (Y12) were purchased from Brilliant Matters. 
Silver ink ECI 1011 LOCTITE (sheet resistance $0.0029\,\Omega/\square$) was obtained from Henkel.
All chemicals and solvents were used as received without further purification.

\subsection{Device fabrication}

\textbf{Solutions}. 
 Branched polyethylenimine was dissolved in ethanol to achieve a concentration of 1\,mg\,ml$^{-1}$. 60\,ml of water and 75\,mg of zinc acetate dihydrate were added to 1\,ml of the ethanol-based PEI solution and ultrasonicated at 40\,°C for 15\,min to dissolve the zinc acetate. A clear and colourless PEI-Zn solution was obtained. The final PEI-Zn solution was stirred at 40\,°C for at least 30\,min and used warm for the printing and spin-coating process.

\textbf{Ink preparation}. 
The active material blends PM6:Y12 (1:1.2 weight ratio) were prepared in air by dissolving the components in either chloroform or o-xylene at the desired total concentrations. For chloroform-based inks, a concentration of 16\,mg\,ml$^{-1}$ was used. For o-xylene-based inks, concentrations of 16\,mg\,ml$^{-1}$, 18\,mg\,ml$^{-1}$ and 22\,mg\,ml$^{-1}$ were prepared. All solutions were stirred overnight at room temperature. The solution (BM-HTL-1) for the hole transport layer was ultrasonicated for 15\,min and filtered using a 0.45\,\textmu{}m PTFE filter.

\textbf{Solar cells}. 
 Cleaned IMI-PET substrates were treated with ozone for 3\,min to increase wettability. Gravure printing of the ETL was conducted immediately after the ozone treatment. Gravure printing was performed using a custom-built proof-printing setup developed at TU Chemnitz. The ethanol-based PEI-Zn solution was gravure-printed under ambient conditions using a printing form with a cell volume of 14\,ml\,m$^{-2}$ at a printing speed of 0.2\,m\,s$^{-2}$ and annealed at 140\,°C for 10\,min in air. The active material blend was immediately gravure-printed onto the annealed ETL using a printing form with a cell volume of 22\,ml\,m$^{-2}$ at a printing speed of 0.2\,m\,s$^{-1}$ and annealed in air at 120\,°C for 5\,min. The BM-HTL-1 solution was slot-die coated under ambient conditions on a preheated stage (60\,°C) at a coating speed of 10\,mm\,s$^{-1}$ using a lab-scale slot-die coater L2005A2-50 from Ossila. The slot-die-coated layers were annealed in air at 120\,°C for 10\,min. Screen printing was carried out using a flatbed screen printer (EKRA). PEDOT:PSS ink was screen-printed (120-34 mesh) on the slot-die-coated HTL and annealed at 130\,°C for 10\,min in air. The silver ink was screen-printed (100-40 mesh) on the PEDOT:PSS layer and annealed at 135\,°C for 2\,min in air. Freshly prepared solar cells were transferred to a nitrogen-filled glovebox and stored in the dark for 48 hours before current--voltage measurements. The prepared device consists of six pixels, each with an active area of 0.2\,cm$^2$.

 Spin-coated reference solar cells were prepared on pre-patterned indium tin oxide (ITO)-coated glass substrates. The patterned substrates were cleaned in detergent, washed with deionised water, acetone and ethanol. The prepared PEI-Zn solution was spin-coated at 4000\,rpm and annealed at 140°C for 10\,min in air. The coated samples were transferred into a nitrogen-filled glovebox. The active material blends were prepared under a protective atmosphere by dissolving PM6:Y12 (1:1.2) in o-xylene and chloroform. The prepared blends were stirred for 24 hours at room temperature and used preheated (40\,°C) for the spin-coating process. The active layers were spin-coated to achieve a thickness of approximately 88\,nm. The BM-HTL-1 solution was spin-coated and annealed at 100\,°C for 5\,min. The silver top electrodes were thermally evaporated. The prepared device consists of six pixels, each with an active area of 0.04\,cm$^2$.

\section{Experimental methods}

\subsection{Current density--voltage measurements}

Current density--voltage (JV) measurements were carried out under an inert nitrogen atmosphere inside a glovebox. The JV characteristics were acquired using a Keithley~2450 source-measure unit, both in the dark and under simulated solar illumination. Illumination equivalent to 1\,sun (100\,mW\,cm$^{-2}$, AM1.5G) was provided by a Wavelabs~LS2 solar simulator.

\subsection{External quantum efficiency}

The external quantum efficiency (EQE) of the devices was measured using wavelength-resolved monochromatic illumination provided by a Bentham monochromator from 280 to 1050\,nm with a step size of 5\,nm. The incident light was modulated with an optical chopper, and the resulting photocurrent response of the device was recorded using a lock-in amplifier (Stanford Research Systems SR830) referenced to the chopper frequency. To correct for the spectral intensity of the light source and to obtain absolute EQE values, the photocurrent signal was normalised to that measured simultaneously with a calibrated Thorlabs FDS1010 silicon photodiode under identical experimental conditions.

\subsection{Sensitive external quantum efficiency}

For sensitive EQE measurements, monochromatic light was generated using an MSHD-300 double monochromator (LOT Quantum Design) coupled to a 100\,W quartz tungsten--halogen lamp. The resulting photocurrent from the devices was detected using a Zurich Instruments MFLI lock-in amplifier. 

The incident light was modulated at a frequency of 223\,Hz using a mechanical chopper (Thorlabs MC2000B-EC). All measurements were carried out with the spectral bandwidth set to 20\,nm. To minimize stray-light artefacts, a series of OD4 long-pass filters with progressively increasing cut-on wavelengths was inserted into the beam path. After spectral filtering, the light was guided through a liquid light guide (Newport 77638) and focused onto the device using a lens. An additional filter (OD4 950\,nm) was inserted at $\lambda=1000$\,nm to reduce stray light artefacts.

Bias voltages were applied to the devices using the voltage output of the Zurich Instruments lock-in amplifier.

The measured photocurrent was normalised to the incident photon flux at each wavelength. The photon flux was determined using a calibrated Hamamatsu K1718-B dual-color photodiode comprising silicon and InGaAs detectors, with the corresponding signals amplified by Thorlabs AMP120 transimpedance amplifiers.

All measurements were performed with the devices mounted in a nitrogen atmosphere inside a Linkam Scientific LTS420 cryostat. The resulting sensitive EQE spectra were finally scaled using the EQE response of a calibrated Thorlabs FDS1010 silicon photodiode.

\subsection{Light-intensity-dependent JV and intensity-modulated photocurrent spectroscopy}

A home-built setup was used to perform light-intensity-dependent JV and $\Voc$ measurements, as well as an intensity-modulated photocurrent spectroscopy (IMPS) measurement. The sample was mounted in a Linkam Scientific LTS420 cryostat via probe-tip electrical contact. The sample temperature was precisely controlled using a Linkam Scientific T96-S temperature controller, which combines liquid-nitrogen cooling and a metal heating stage. A continuous-wave diode laser (Omicron~LDM~A350) with a center wavelength of 515\,nm and fast analog modulation capability up to 350\,MHz was used to provide modulated and continuous illumination. The illumination intensity was controlled using a pair of motorised neutral-density filter wheels (FW102C and FW112C, Thorlabs), which can provide several orders of magnitude illumination intensity variation. The illumination intensity was monitored with a Newport 818-BB-40 biased silicon photodiode. For JV measurement, a Keithley~2634b source measure unit was used to supply voltage bias to the sample as well as measure the current from both the sample and photodiode. For IMPS measurement, the sample is approximately under short-circuit conditions. A Zurich Instruments UHFLI lock-in amplifier was used to apply the modulation signal of the laser and to record the photocurrent of the sample and photodiode. The modulation amplitude was set to 10\,\% of the bias illumination intensity to ensure small-signal excitation.

\subsection{Variable-angle spectroscopic ellipsometry}

Optical modelling of variable-angle spectroscopic ellipsometry (VASE) data was performed to extract the optical constants and absorption coefficients of spin-coated and printed solar cells up to the active layer, including the ITO/glass or IMI-PET substrates and the PEI-Zn electron transport layer. 
VASE measurements were performed at 7 different angles from 45$^\circ$ - 75$^\circ$ in steps of 5$^\circ$ and in the spectral range from 0.7 - 5.0\,eV using the M2000 T-Solar ellipsometer by J.\ A.\ Woollam. The CompleteEASE software by J.\ A.\ Woollam was used to perform optical modelling. 
For the extraction of the optical response of the individual layers, VASE measurements were performed on clean substrates, substrates with only the PEI-Zn transport layer, and finally on substrates with PEI-Zn and the PM6:Y12 active layer.
In order to suppress reflections from the backside of the transparent substrates, they were mechanically roughened. 
For the determination of the optical response of glass in the ITO/glass substrates and of PET in the IMI-PET substrates, the electrode material was mechanically removed, also roughening the surface to suppress reflections, and VASE measurements were performed from the backside. 

The IMI-electrode consists of three layers -- a thin Ag layer in-between two ITO layers. Since the thickness and optical response of neither of the layers were known, certain assumptions were made to determine these parameters: the ITO layers are of equal thickness and constitute the same optical response. For the Ag-layer we assumed the dielectric function by Palik et al.\cite{palik1998handbook} For the modelling of all dielectric functions, a first approximation of film thickness was made using a Cauchy function to fit spectral ranges which are transparent or have low levels of absorption. Thereafter, the absorptive response was approximated using a Kramers-Kronig consistent B-Spline, while keeping the film thickness either fixed or variable within narrow limits. This was followed by a parametrisation of the imaginary part of the dielectric function with general oscillator models, and the calculation of the real part thereof. Surface roughness was accounted for by a Bruggeman effective medium approximation of a 1:1 mixture of void and the surface layer. 

Furthermore, especially for the printed samples, non-idealities had to be accounted for, in terms of thickness non-uniformity and angular spread, due to small waves in the flexible substrate. In the model this is done by a convolution of generated spectra where thickness and surface orientation are varied. The degree of variation is the fitting parameter for these properties. Absorption onsets near the energy gap were modelled using the Cody-Lorentz oscillator, while other absorptive features were modelled with Gaussian oscillators. In the case of the active layers, significant optical anisotropy was accounted for by the conversion of the isotropic general oscillator models into uniaxially anisotropic ones. To avoid over-parametrisation, only the oscillator amplitudes were kept as free parameters in an initial fitting step. Finally, the thickness was fitted together with the oscillator amplitudes and the non-ideality parameters.

\subsection{Steady-state absorption and photoluminescence}

Steady-state absorption and photoluminescence (PL) measurements were performed using an Edinburgh Instruments FLS1000 modular spectrometer. All measurements were carried out at room temperature, with the samples mounted under ambient conditions.

Excitation light was provided by a 450\,W xenon lamp and wavelength was selected using the excitation monochromator. Steady-state absorption spectra were recorded in transmission mode. Thin films were measured on the same substrates as used for device fabrication, with a bare substrate serving as reference. Absorbance spectra were calculated from the transmission data and converted to absorptance where required for further analysis.

For PL measurements, monochromatic excitation was selected with the excitation monochromator, using excitation wavelengths of 500\,nm for blend films and 510\,nm for neat Y12 and PM6 films. The emitted PL was dispersed by the emission double monochromator and detected using a near-infrared photomultiplier tube (nitrogen-flow-cooled housing, spectral range 500--1400\,nm, temporal resolution 800\,ps).

\subsection{Profilometry}

Layer thickness was determined using a stylus profilometer Dektak TX Bruker equipped with a stylus apex radius of 6.5\,\textmu{}m. For thickness measurements, the layers were mechanically scratched to expose the underlying substrate, and the surface profile was obtained by scanning across the scratched region.

\subsection{Optical imaging and laser scanning microscopy}

Images of the printed layers were taken using the optical imaging technique. Transmitted-light scans were recorded using an Epson Perfection V850 Pro scanner and analysed with a custom MATLAB routine. The images were converted to grayscale, and the intensity values were normalised using a dark (opaque) reference and an uncoated-substrate baseline. A spin-coated calibration series with known thicknesses, measured independently by stylus profilometry, was prepared and scanned under identical conditions to establish an intensity--thickness calibration curve. This calibration was then applied pixel-wise to the scanned samples to generate thickness maps. For each sample, thickness statistics and vertical thickness profiles were extracted from fixed-size regions of interest. 

Magnified images were obtained using a laser scanning microscope (Keyence~VK-9700). No additional image analysis was performed.

\subsection{Contact angle measurements}

Blends with different PM6:Y12 ratios were prepared at the desired weight ratios and spin-coated onto pre-cleaned glass substrates to obtain uniform thin films. Contact angle measurements were performed on OCA 20 DataPhysics using the sessile drop method with three probe liquids (water, diiodomethane and ethylene glycol). For the PEI-Zn electron transport layer, surface energy calculations were performed using glycerol, 1-bromonaphthalene, and diiodomethane as probe liquids. The surface energy was calculated using the Owens--Wendt--Rabel--Kaelble (OWRK) method.

\clearpage
\section{Supplementary data for printing process optimisation}

\begin{table}[h]
    \centering
    \caption{Median layer thickness and uniformity of gravure-printed o-xylene-based PM6:Y12 films. The data were obtained from transmitted-light scans over an area of $10\,\mathrm{mm} \times 10\,\mathrm{mm}$.}
    \begin{tabular}{c|c|c|c}
        \hline
        Blend & Gravure form  & Layer & Coefficient \\
        concentration & cell volume & thickness [nm] & of variation [\%] \\
        \hline
        22\,mg\,ml$^{-1}$ & 22\,ml\,m$^{-2}$ & $102 \pm 10$ & 9.0 \\
        18\,mg\,ml$^{-1}$ & 22\,ml\,m$^{-2}$ & $88 \pm 3$  & 2.1 \\
        16\,mg\,ml$^{-1}$ & 22\,ml\,m$^{-2}$ & $67 \pm 3$  & 3.7 \\
        18\,mg\,ml$^{-1}$ & 26\,ml\,m$^{-2}$ & $128 \pm 3$  & 3.9 \\
        \hline
    \end{tabular}
    \label{SI_tab:printed_uniformity}
\end{table}

\begin{figure}[h]
    \centering
    \includegraphics[width=0.45\linewidth]{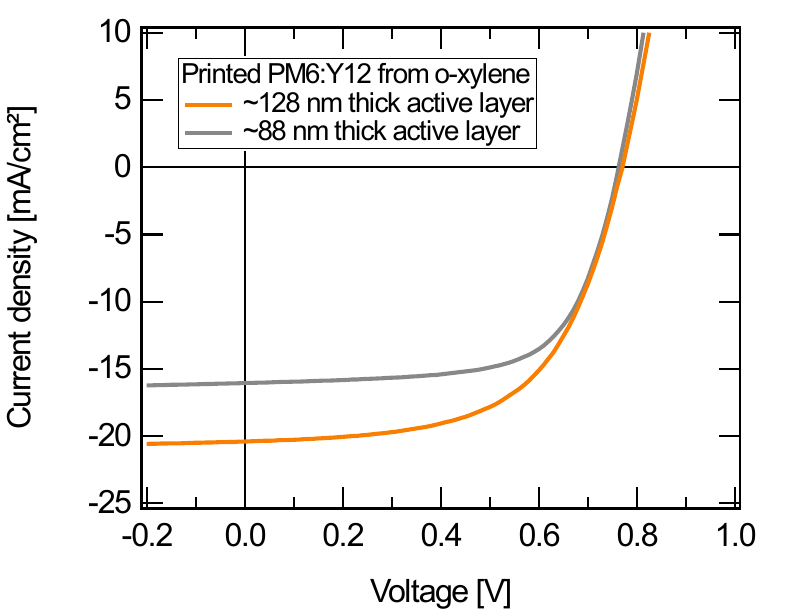}
    \caption{Current--voltage characteristics of printed PM6:Y12 solar cells based on thick ($\sim$128\,nm) and thin ($\sim$88\,nm) active layers. Even though thick active layers lead to increased $\jsc$, the study focuses on thin active-layer solar cells for better comparison to spin-coated references.}
    \label{SI_fig:JV_thick_slim}
\end{figure}

Current--voltage characteristics of thick and thin active-layer printed solar cells are shown in Figure~\ref{SI_fig:JV_thick_slim}.

\begin{figure}[h]
    \centering
    \includegraphics[width=0.9\linewidth]{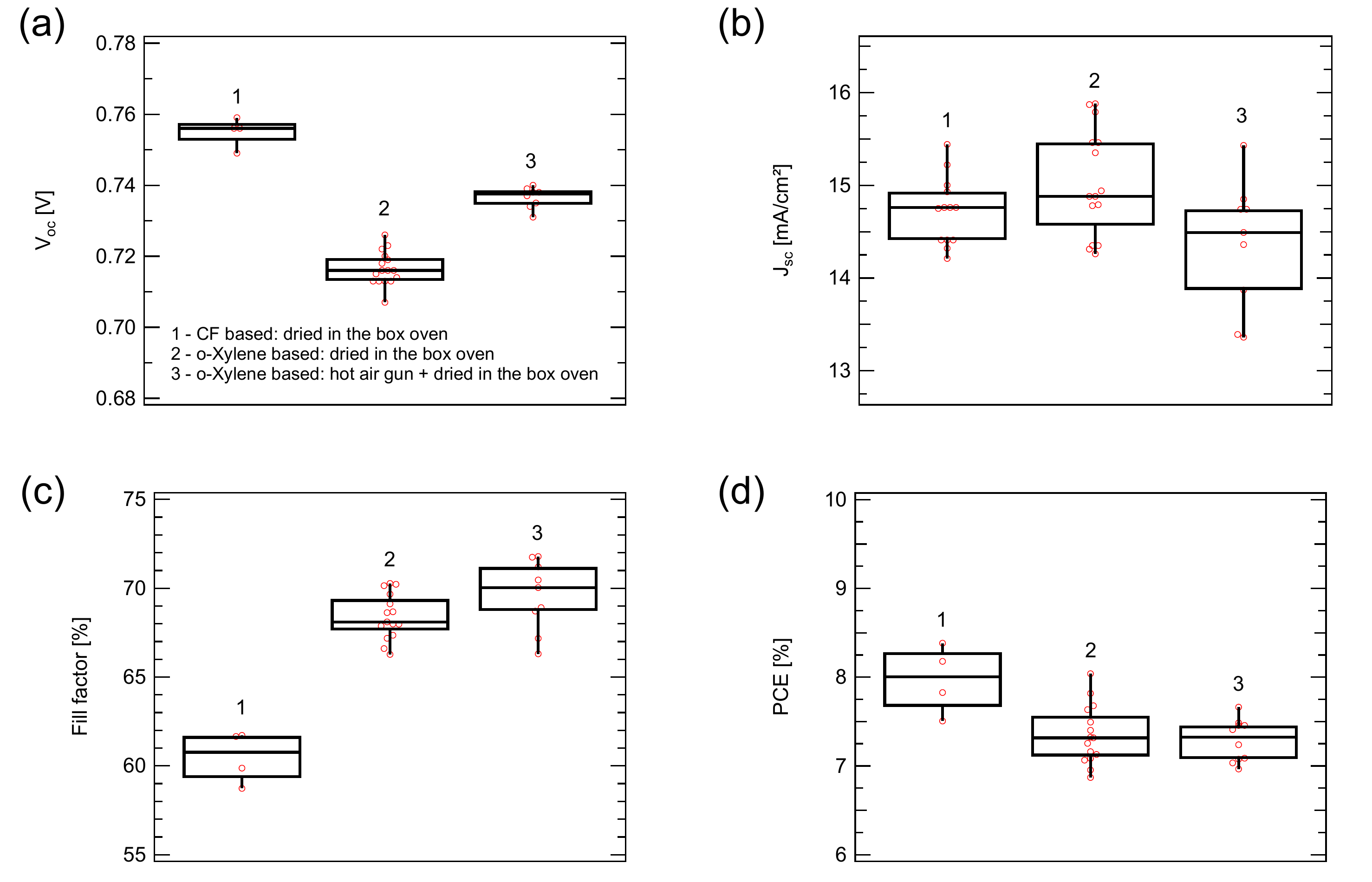}
    \caption{Statistics of the solar cell characteristics for different drying conditions. $\jsc$ and PCE are given as measured on Wavelabs~LS2 solar simulator, i.e.\ uncorrected for the solar simulator spectrum mismatch and PEDOT:PSS area. }
    \label{SI_fig:drying}
\end{figure}

The effect of drying conditions on the solar cell characteristics is shown in Figure~\ref{SI_fig:drying}. Drying conditions during proof-printing have minimal impact on the performance of chloroform- and o-xylene-based cells. 
Devices processed using a combined air-dryer and box-oven drying sequence exhibited slightly improved electrical characteristics compared to samples dried in the box-oven only, with increases observed in $\Voc$ and $\FF$. However, box-oven drying alone was found to be sufficient to achieve effective solvent removal for both chloroform- and o-xylene-based blends without compromising device performance. Consequently, a one-step box-oven drying protocol at 120\,°C for 5\,min was adopted as the standard processing condition for printed devices.

\begin{figure}[h]
    \centering
    \includegraphics[width=0.9\linewidth]{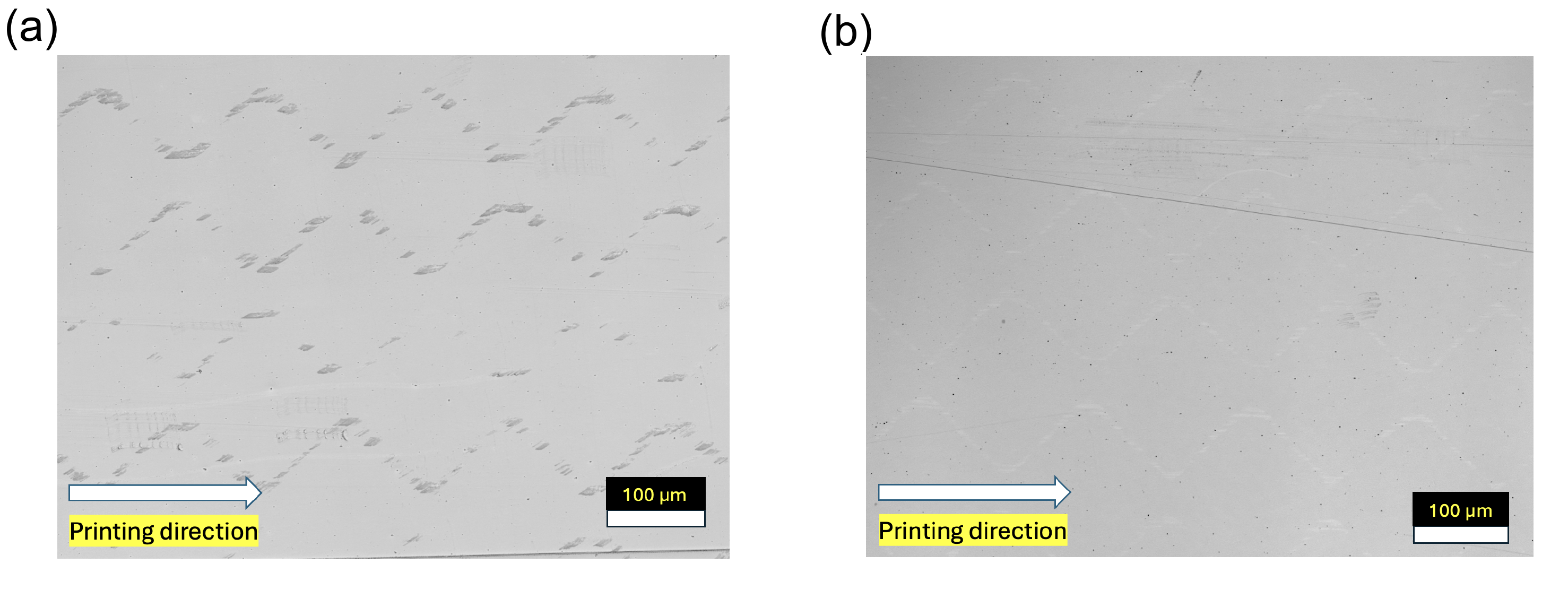}
    \caption{Optical images illustrating the mechanical stability of (a) gravure-printed nano-ZnO layer, and (b) gravure-printed PEI-Zn layers under identical testing conditions.}
    \label{SI_fig:ETL_gravure}
\end{figure}


\begin{figure}[h]
    \centering
    \includegraphics[width=0.5\linewidth]{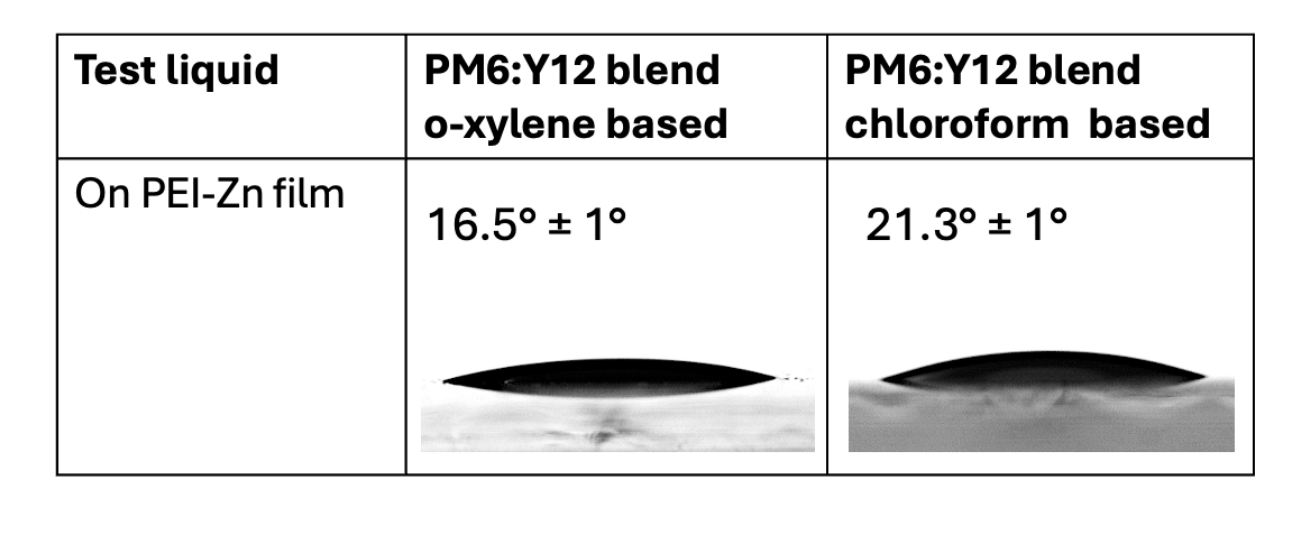}
    \caption{Contact-angle measurements showing wettability of PM6:Y12 blends on a spin-coated PEI-Zn electron transport layer.}
    \label{SI_fig:AL_wetting}
\end{figure}


\begin{figure}[h]
    \centering
    \includegraphics[width=0.9\linewidth]{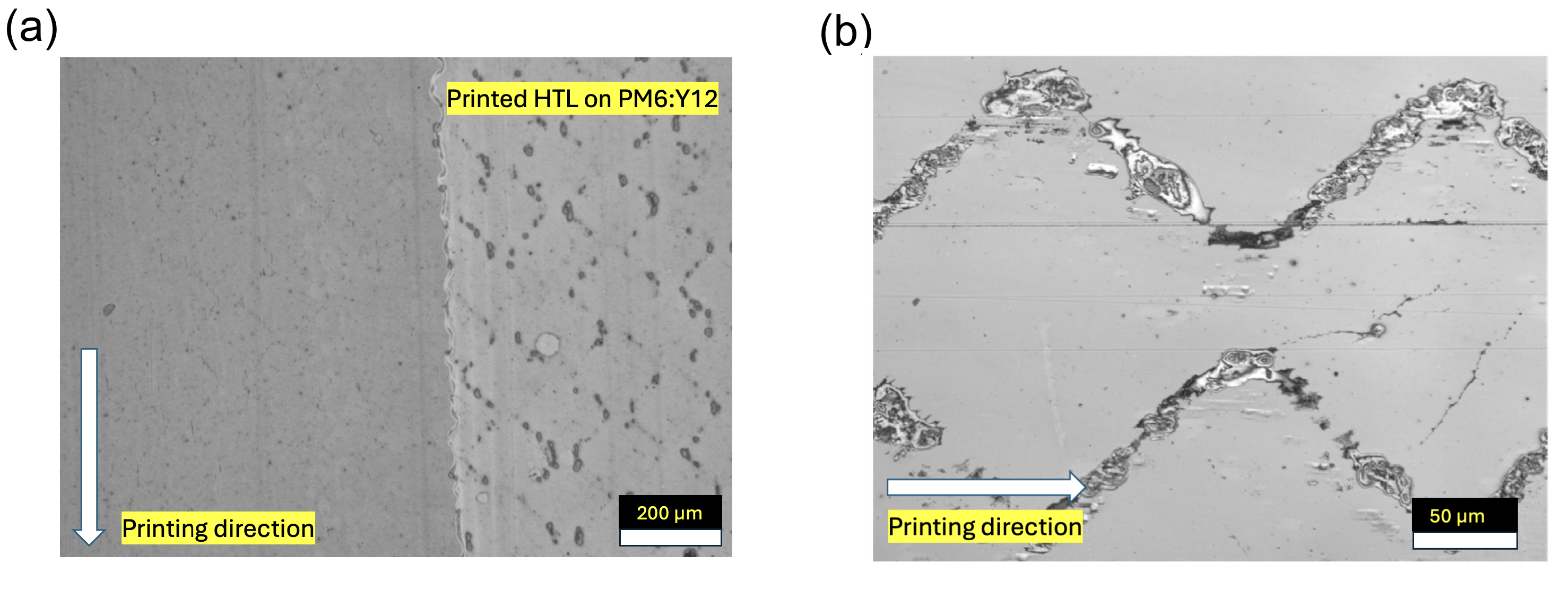}
    \caption{Laser scanning microscope images illustrating mechanical damage of the PM6:Y12 active layer induced by printing of the hole-transport layer. (a) Overview image showing impressions of gravure printing form after HTL printing. The arrow indicates the printing direction. (b) Representative magnified view of a damaged region, highlighting local tearing of the active layer. Images are shown in their original orientation}
    \label{SI_fig:HTL_gravure}
\end{figure}

\clearpage
\section{Supplementary JV and EQE data}

\begin{figure}[h]
    \centering
    \includegraphics[width=0.9\linewidth]{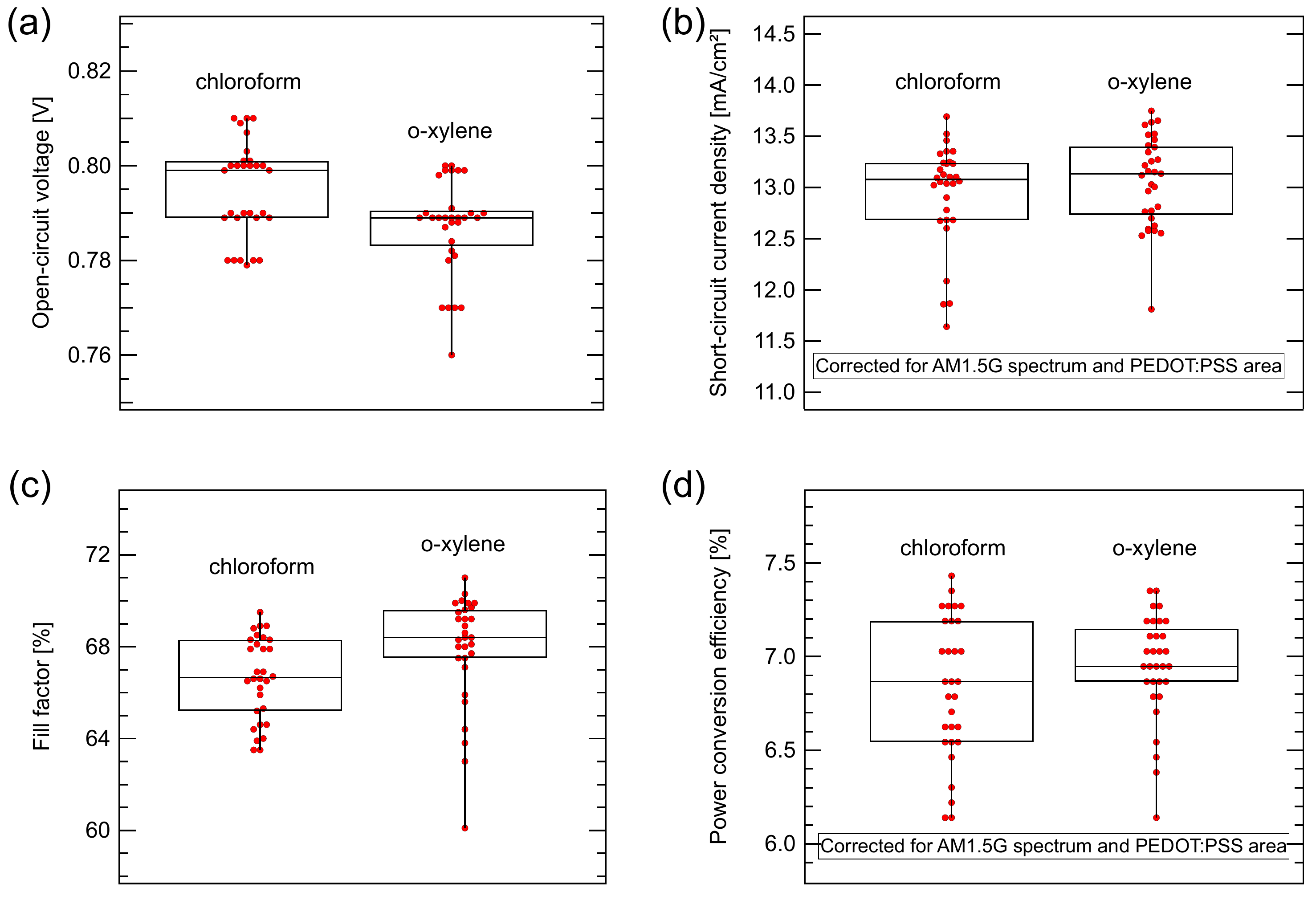}
    \caption{Solar cell characteristics of chloroform- and o-xylene-based printed solar cells: a) open-circuit voltage ($\Voc$), b) short-circuit current density ($\jsc$), c) fill factor ($\FF$), and d) power conversion efficiency (PCE). The data for $\jsc$ was measured using Wavelabs~LS2 solar simulator and then corrected for AM1.5G spectrum and the PEDOT:PSS area. The power conversion efficiency was corrected accordingly.}
    \label{SI_fig:JVstats}
\end{figure}

\begin{table}[h]
\caption{Solar cell parameters for gravure-printed devices based on chloroform and o-xylene, as shown in Figure~\ref{SI_fig:JVstats}. The values were obtained from 30 devices for chloroform-based and 31 devices for o-xylene-based solar cells, and are presented as mean $\pm$ standard deviation. The values of $\jsc^\mathrm{*}$ and PCE$^\mathrm{*}$ have been corrected for the AM1.5G spectrum and PEDOT:PSS area. Slight differences in PCE values originate from rounding.}
\centering
\small
\renewcommand{\arraystretch}{1.15}
\begin{tabular*}{0.9\linewidth}{@{\extracolsep{\fill}}lccccc@{}}
\hline
Device type & $\Voc$ & $\jsc$ & $\jsc^\mathrm{*}$ & $FF$ & PCE$^\mathrm{*}$ \\
& (mV) & (mA\,cm$^{-2}$) & (mA\,cm$^{-2}$) & (\%)  & (\%) \\
\hline
chloroform  & $795 \pm 10$ & $16.02 \pm 0.62$ & $12.94 \pm 0.50$ & $66.64 \pm 1.80$ & $6.84 \pm 0.38$ \\
o-xylene & $787 \pm 10$ & $16.17 \pm 0.55$ & $13.06 \pm 0.44$ & $67.89 \pm 2.43$ & $6.95 \pm 0.28$ \\
\hline
\multicolumn{6}{l}{\footnotesize $^\ast$ AM1.5G and area-corrected $\jsc$ and PCE values.}
\end{tabular*}\label{SI_tab:PCEstats}
\end{table}

\begin{figure}[h]
    \centering
    \includegraphics[width=0.9\linewidth]{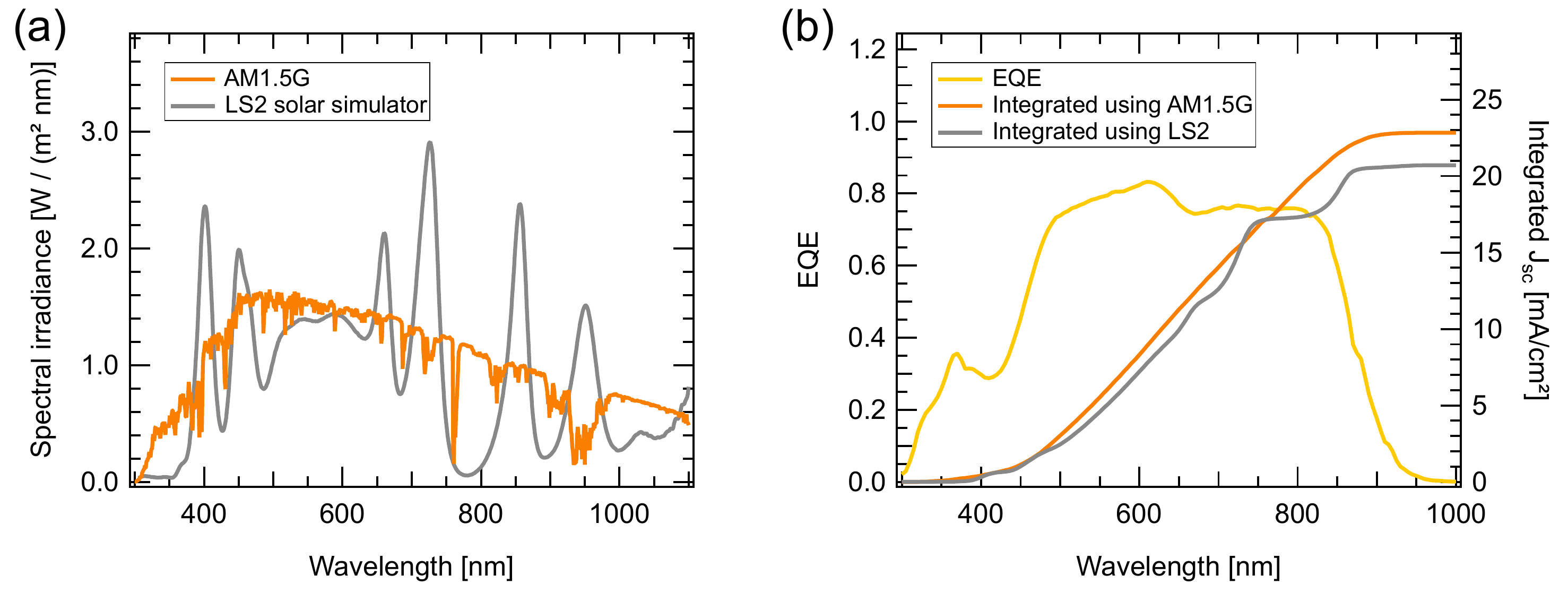}
    \caption{(a) Emission spectrum of the Wavelabs~LS2 solar simulator in comparison with the AM1.5G reference spectrum. (b) Short-circuit current density obtained by integrating the EQE with the LS2 and AM1.5G spectra, indicating a $\jsc$ reduction of $\sim$10\,\% due to spectral mismatch.}
    \label{SI_fig:spectrum}
\end{figure}
\vspace{2cm}

\begin{figure}[h]
    \centering
    \includegraphics[width=0.9\linewidth]{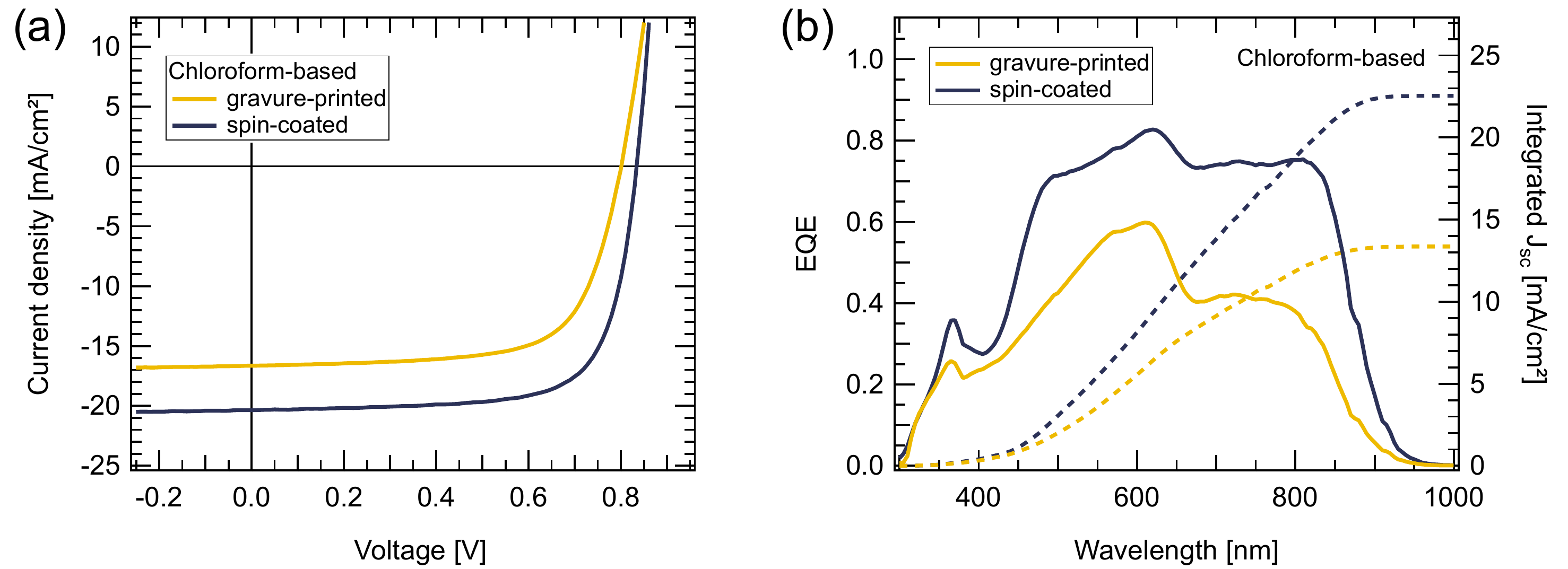}
    \caption{(a) JV characteristics, and (b) EQE spectra of gravure-printed and spin-coated chloroform-based PM6:Y12 devices. JV curves were measured under 1\,sun equivalent illumination provided by the Wavelabs~LS2 solar simulator (for spectrum, see Figure~\ref{SI_fig:spectrum}).}
    \label{SI_fig:CF_JV_EQE}
\end{figure}

\begin{table}[h]
\caption{Literature overview of PCE values for R2R-compatible OSCs, including slot-die coating, inkjet printing, and gravure printing of the AL. The table lists the AL system, material type (fullerene or non-fullerene acceptor), deposition method, reported PCE, publication year, and corresponding reference. The dataset corresponds to the values shown in Figure~\ref{fig:JV_EQE}(c).}
\centering
\small
\renewcommand{\arraystretch}{1.15}
\begin{tabular*}{0.9\linewidth}{@{\extracolsep{\fill}}lccccc@{}}
        \hline
        Active-layer system & Type & Method & PCE (\%) & Year & Ref.\ \\
        \hline
        P3HT:PCBM       & fullerene     & gravure   & 1.30   & 2011 & \citenum{huebler2011printed}\\
        P3HT:PCBM       & fullerene     & slot-die  & 1.82  & 2012 & \citenum{yu2012silver}\\
        PSBTBT:PDI-DTT  & non-fullerene & slot-die  & 0.20   & 2013 & \citenum{liu2013all}\\
        PDTSTTz-4:PCBM  & fullerene     & slot-die  & 3.50   & 2013 & \citenum{helgesen2013slot}\\
        P3HT:PCBM       & fullerene     & slot-die  & 1.60   & 2013 & \citenum{hoesel2013fast}\\
        PBDTTT-C-T:PCBM & fullerene     & slot-die  & 2.09  & 2014 & \citenum{cheng2014comparison}\\
        PCDTBT:PCBM     & fullerene     & inkjet    & 2.05  & 2014 & \citenum{Jung2014}\\
        P3HT:PCBM       & fullerene     & inkjet    & 1.70   & 2015 & \citenum{eggenhuisen2015high}\\
        PV2000:PCBM     & fullerene     & inkjet    & 4.10   & 2015 & \citenum{eggenhuisen2015high}\\
        P3HT:PCBM       & fullerene     & slot-die  & 1.15  & 2015 & \citenum{galagan2015roll}\\
        P3HT:PCBM       & fullerene     & gravure   & 2.90   & 2015 & \citenum{vilkman2015gravure}\\
        PTB7-Th:IEIC    & non-fullerene & slot-die  & 2.26  & 2016 & \citenum{liu2016roll-coating}\\
        P3HT:PCBM       & fullerene     & gravure   & 2.22  & 2016 & \citenum{kapnopoulos2016fully}\\
        P3HT:PCBM       & fullerene     & gravure   & 1.50   & 2016 & \citenum{vak2016reverse}\\
        low $\Eg$ copolymer:PCBM & fullerene& slot-die & 6.00     & 2021 & \citenum{miranda2021efficient}\\
        PV2000:PCBM     & fullerene     & inkjet    & 3.00     & 2024 & \citenum{Steinberger2024}\\
        P3HT:o-IDTBR    & non-fullerene & slot-die  & 4.70   & 2025 & \citenum{feroze2025long}\\
        PM6:Y7-12       & non-fullerene & slot-die  & 11.51  & 2025 & \citenum{jayaraman2025flexible}\\
        PM6:Y12         & non-fullerene & gravure   & 7.32  & 2026 & this work\\
        \hline
\end{tabular*}\label{SI_tab:PCE_lit}
\end{table}

\clearpage
\section{Optical modelling}\label{SI_sec:optical_sim}

\begin{figure}[h]
    \centering
    \includegraphics[width=0.9\linewidth]{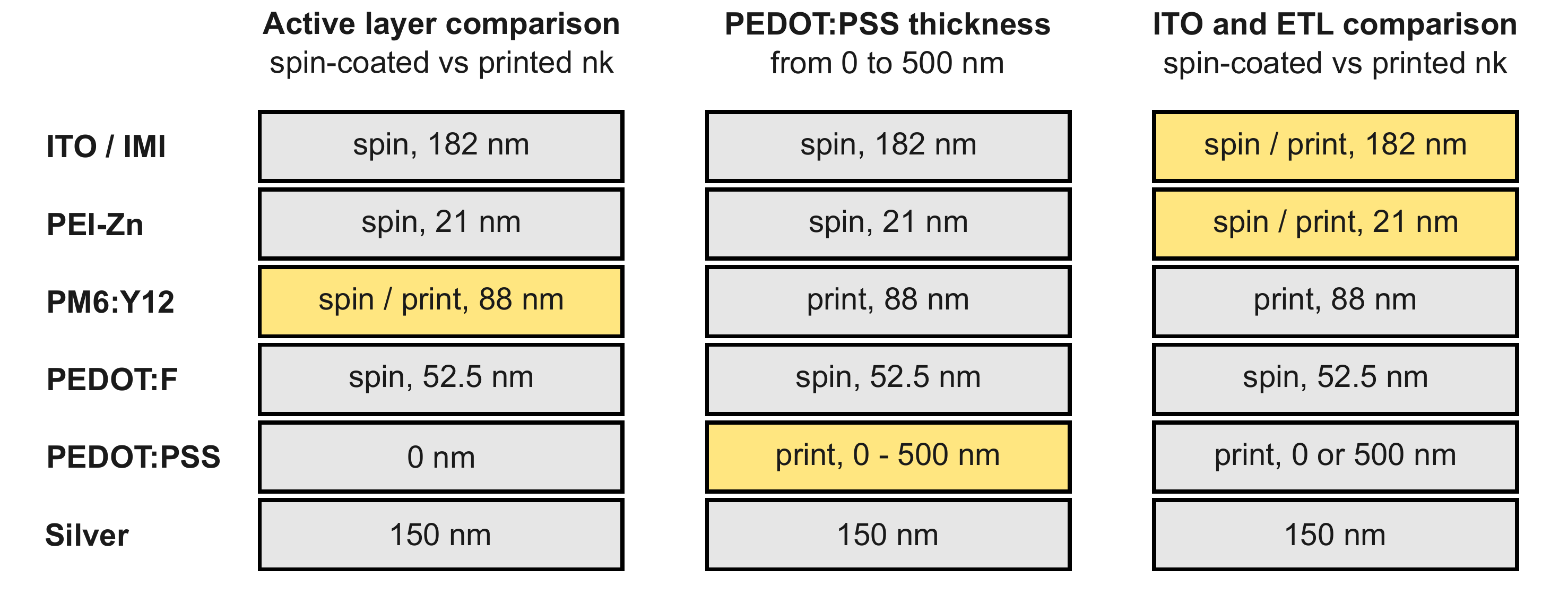}
    \caption{Layer stacks used for the optical modelling: (left) to isolate the effect of the active-layer optical constants, where only the active-layer $n,k$ are varied; (middle) to evaluate the influence of the PEDOT:PSS layer thickness; (right) with different bottom electrode and ETL configurations. Here, \emph{print} and \emph{spin} indicate that the optical constants ($n,k$) were measured on printed and spin-coated stacks, respectively.}
    \label{SI_fig:transfermx}
\end{figure}

Optical generation profiles were obtained using simulations performed with the OghmaNano software. The simulations were carried out under AM1.5G illumination (1000\,W\,m$^{-2}$) and used experimentally determined optical constants (refractive index and extinction coefficient obtained from VASE measurements) as input. The optical stack was discretised using a spatial mesh with a step size of 2\,nm, and the optical response was calculated over a wavelength range from 300 to 1200\,nm using 226 discrete wavelength points. From this, the spatially resolved optical generation rate was obtained for the complete device stack. The layer sequences and thicknesses employed in the simulations are shown in Figure~\ref{SI_fig:transfermx}.

\clearpage
\section{Determination of the exciton quenching ratio}\label{SI_sec:PL}

\begin{figure}[h]
    \centering
    \includegraphics[width=0.85\linewidth]{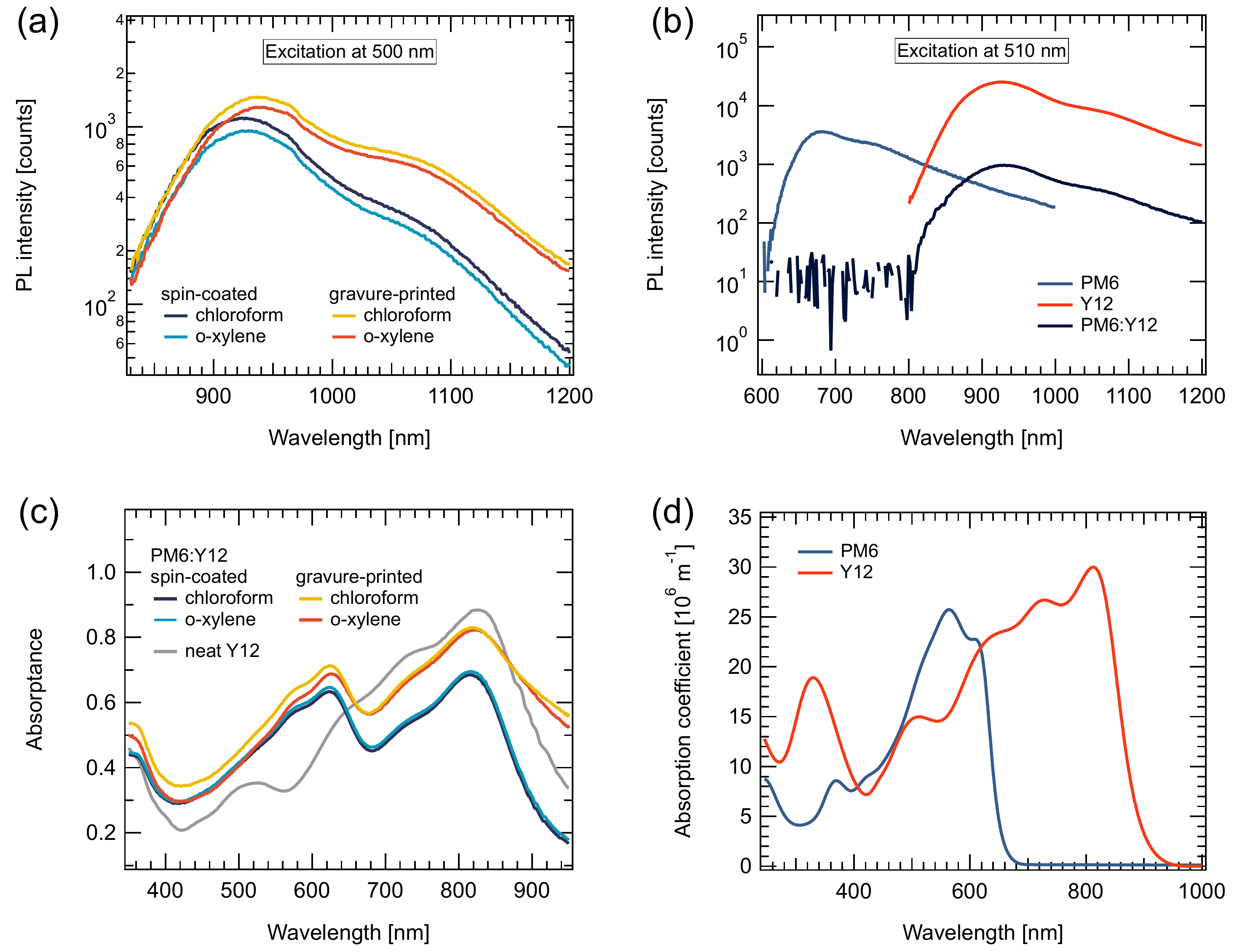}
    \caption{PL spectra of (a) PM6:Y12 blend films under excitation at 500\,nm for spin-coated and printed samples processed from chloroform and o-xylene; (b) neat PM6, neat Y12, and PM6:Y12 blend films under excitation at 510\,nm. (c) Absorptance spectra of the blend films and neat Y12 used to normalize the integrated PL intensity at the excitation wavelength. (d) Absorption coefficients of PM6 and Y12 determined by ellipsometry.}
    \label{SI_fig:PL}
\end{figure}

Photoluminescence (PL) measurements were used to quantify exciton quenching in PM6:Y12 blends. The PL signal shown in Figure~\ref{SI_fig:PL}(a) was integrated over the wavelength range from 830 to 1200\,nm, where emission from Y12 dominates (Figure~\ref{SI_fig:PL}(b)).

The integrated PL intensity was normalised by the optical absorptance at the excitation wavelength (500\,nm for blends and 510\,nm for neat Y12). The absorptance spectra are shown in Figure~\ref{SI_fig:PL}(c). This normalisation yields a PL intensity per absorbed photon and removes first-order effects related to differences in film thickness or absorption strength.

The exciton quenching efficiency $\etaexc$ discussed in the main text was calculated as
\begin{equation}
\etaexc = 1 - \frac{(\mathrm{PL}/A)_{\mathrm{blend}}}{(\mathrm{PL}/A)_{\mathrm{Y12}}} , 
\end{equation}
where $(\mathrm{PL}/A)_{\mathrm{blend}}$ and $(\mathrm{PL}/A)_{\mathrm{Y12}}$ denote the absorptance-normalised integrated PL intensities of the blend and neat Y12 films, respectively.

Figure~\ref{SI_fig:PL}(d) shows the absorption coefficients of PM6 and Y12 determined for the ellipsometry measurements. Both PM6 and Y12 absorb at 500\,nm. However, we assume that Y12 emission under 500\,nm excitation is predominantly fed by excitons generated in the blend as a whole, including excitons initially created on the donor and transferred to Y12. Therefore, normalisation by the Y12-only absorptance in the blend was not applied, and the total blend absorptance at the excitation wavelength was used instead.

\clearpage
\section{Absorbance spectra}

\begin{figure}[h]
    \centering
    \includegraphics[width=0.95\linewidth]{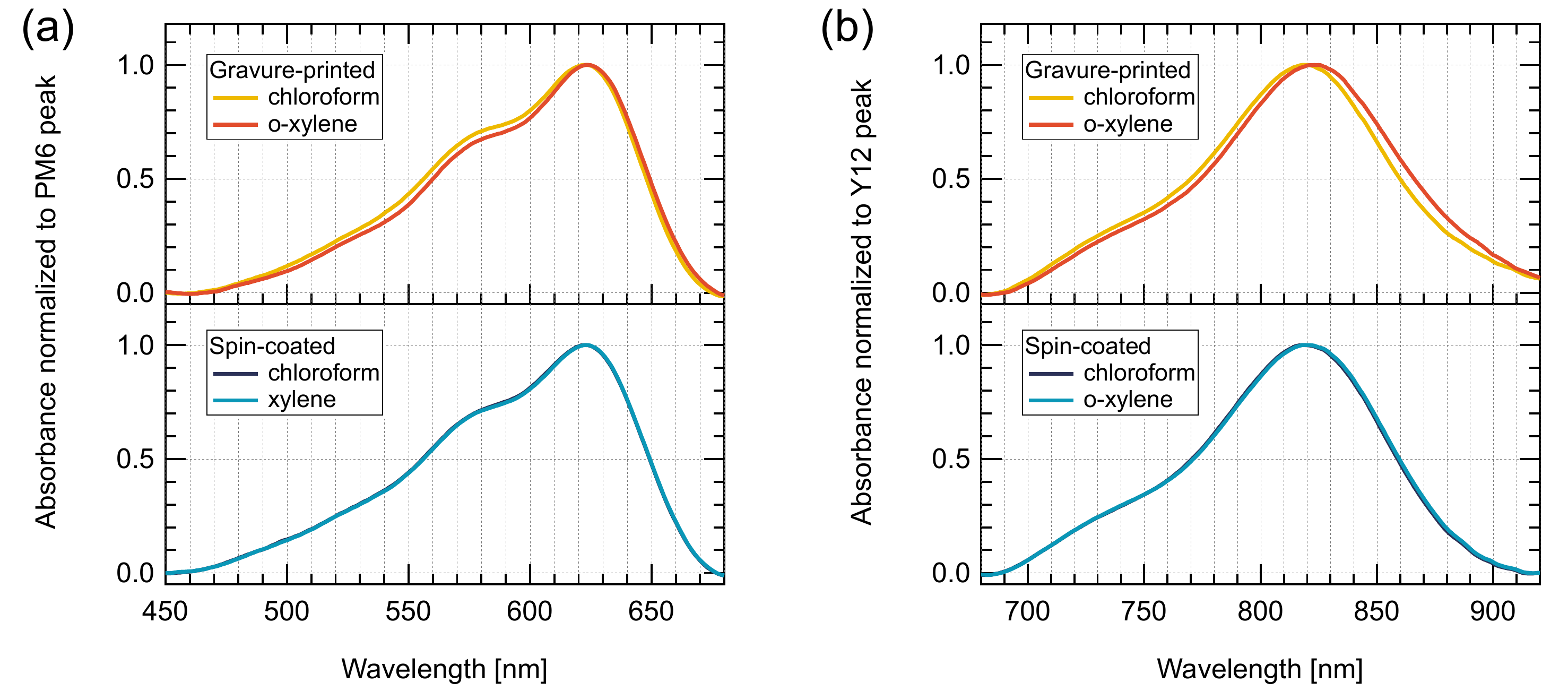}
    \caption{Absorbance spectra of PM6:Y12 blend films (a) normalised to PM6 peak, and (b) normalised to Y12 peak.}
    \label{SI_fig:abs}
\end{figure}

Absorbance spectra of PM6:Y12 blend films were normalised to the absorption peaks of PM6 and Y12 in order to study the spectral shifts (Figure~\ref{SI_fig:abs}). Prior to normalisation, baseline correction was performed for the absorbance spectra in MATLAB using an asymmetric least-squares (AsLS) algorithm with second derivative smoothness regularisation.\cite{He2014AsLS}

\section{Surface energy determination}\label{SI_sec:surfaceE}

\begin{figure}[h]
    \centering
    \includegraphics[width=0.9\linewidth]{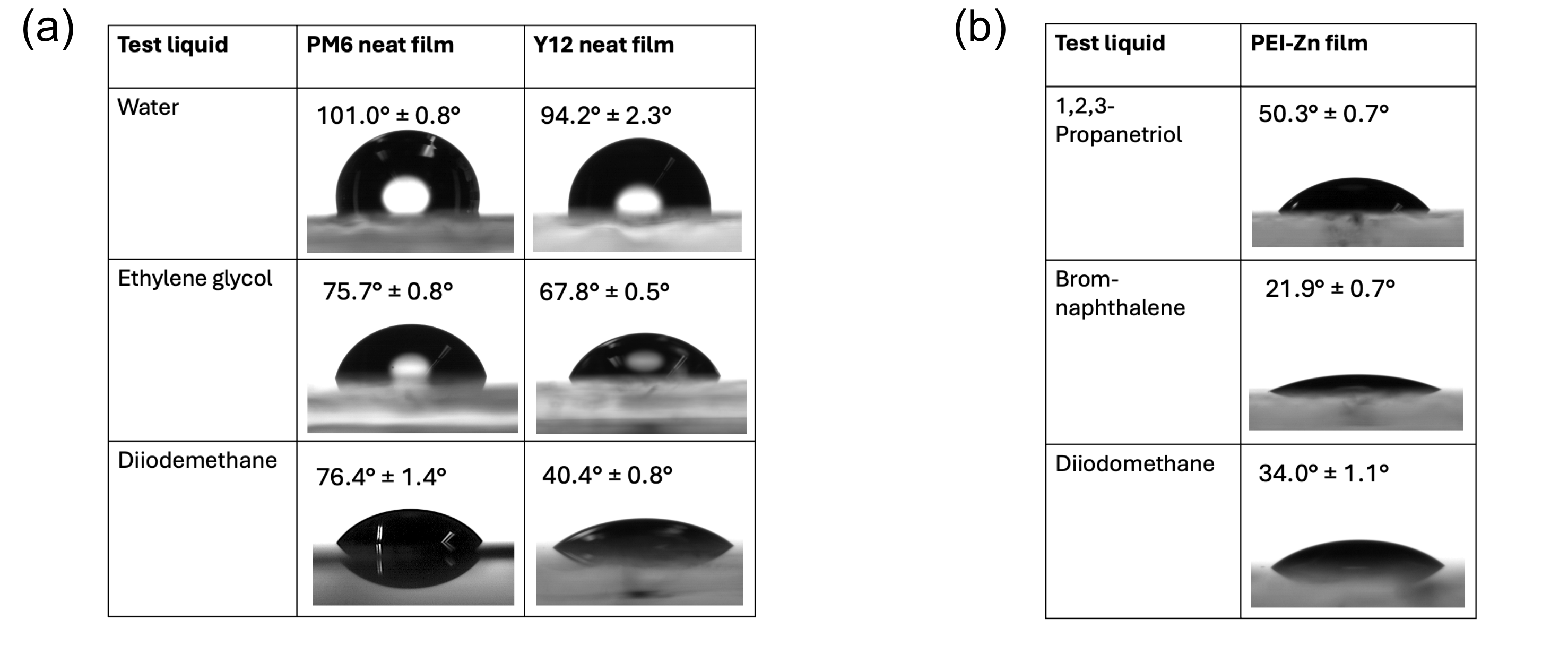}
    \caption{Contact-angle measurements showing wetting behaviour of test liquids. (a) On neat spin-coated PM6 and Y12 films. (b) =n neat spin-coated PEI-Zn films.}
    \label{SI_fig:contact_angles}
\end{figure}

To calculate the surface energy, we use the Owens--Wendt--Rabel--Kaelble (OWRK) method,\cite{owens1969estimation} where the total surface energy of the solid ($\gamma_{S}$) is expressed as the sum of its dispersive ($\gamma_{S}^{d}$) and polar ($\gamma_{S}^{p}$) components:
$\gamma_{S} = \gamma_{S}^{d} + \gamma_{S}^{p}$. According to the OWRK model, the contact angle $\theta$ satisfies 
$$\gamma_{L}\left(1+\cos\theta\right)=2\left(\sqrt{\gamma_{S}^{d}\gamma_{L}^{d}}+\sqrt{\gamma_{S}^{p}\gamma_{L}^{p}}\right) , $$
where $\gamma_{L}$ is the total surface tension of the probe liquid, and $\gamma_{L}^{d}$ and $\gamma_{L}^{p}$ are its dispersive and polar components, respectively. The parameters $\gamma_{S}^{d}$ and $\gamma_{S}^{p}$ were determined from the measured contact angles using the known surface tension components of the probe liquids, as shown in Figure~\ref{SI_fig:contact_angles}. The surface energies of neat Y12 and PM6 (Figure~\ref{SI_fig:contact_angles}(a))) are $\gamma_{\mathrm{Y12}} = 37.15 \pm 0.41\,\mathrm{mN\,m^{-1}}$ and $\gamma_{\mathrm{PM6}} = 30.20 \pm 1.75\,\mathrm{mN\,m^{-1}}$, respectively. The surface energy of PEI-Zn layer (49.10\,mN\,m$^{-1}$, Figure~\ref{SI_fig:contact_angles}(b)) is substantially higher than that of PM6, making polymer enrichment at the ETL interface energetically unfavourable. 

\begin{figure}[h]
    \centering
    \includegraphics[width=0.45\linewidth]{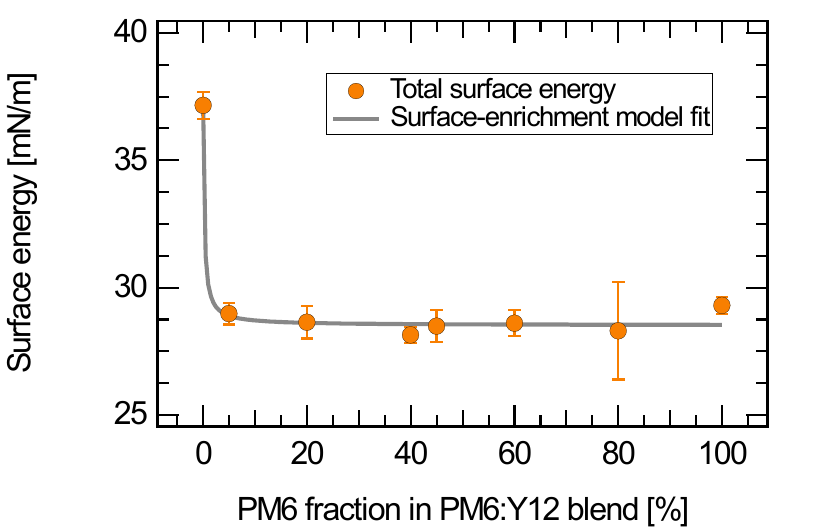}
    \caption{Total surface energy of PM6:Y12 blends with different PM6 fractions and fit using the surface enrichment model.}
    \label{SI_fig:surfaceE}
\end{figure}

For spin-coated films, the surface energies in Figure~\ref{SI_fig:surfaceE} remain close to those of neat PM6 over a wide range of blend compositions, indicating preferential enrichment of PM6 at the air interface. Only at low PM6 fractions does the surface composition transition toward Y12-rich, indicating a change in the dominant surface species. 

We analysed the total surface energy data using a thermodynamic surface-enrichment model with a Langmuir-type functional form, in which the surface composition differs from the bulk due to differences in interfacial free energy between the blend components.\cite{Jones1989PRL} The model assumes equilibrium partitioning between the bulk and the surface, characterised by an enrichment factor 
\[
K = \exp\left(-\frac{\Delta G_{\mathrm{seg}}}{k_{\mathrm{B}} T}\right),
\]
analogous to the Langmuir adsorption isotherm. Here, $\Delta G_{\mathrm{seg}}$ denotes the Gibbs free energy of segregation, i.e., the free energy difference between a component located at the surface and in the bulk.
The surface polymer fraction, $\phi_{\mathrm{surf}}(w)$, is related to the bulk polymer fraction $w$ according to 
$$\phi_{\mathrm{surf}}(w) = \frac{K\,w}{1 + (K - 1)\,w} , $$
where $K$ is the surface-enrichment factor representing the relative preference of the polymer for the air interface. The total surface energy is then expressed as
$$\gamma(w) = \gamma_{\mathrm{Y12}} + \left(\gamma_{\mathrm{PM6}} - \gamma_{\mathrm{Y12}}\right) \phi_{\mathrm{surf}}(w) , $$
where $\gamma_{\mathrm{Y12}}$ and $\gamma_{\mathrm{PM6}}$ denote the surface energies of the Y12-rich and PM6-rich phases, respectively. 

The fitting of data in Figure~\ref{SI_fig:surfaceE} yielded an enrichment parameter $\ln K = 1.65 \pm 1.05$ ($n = 8$). The positive value of $\ln K$ indicates preferential enrichment of PM6 at the surface and consistent with interfacial free-energy minimisation.

\clearpage
\section{Determination of open-circuit voltage losses}
\label{SI_sec:Voc}

\begin{figure}[h]
    \centering
    \includegraphics[width=0.45\linewidth]{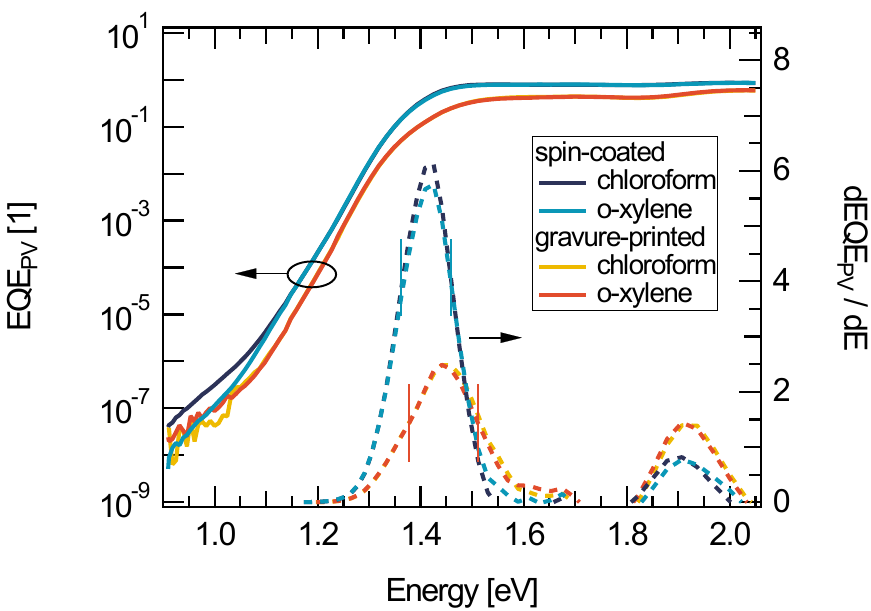}
    \caption{EQE spectra and the optical gap determined from $\mathrm{d\,EQE}_{\mathrm{PV}}/\mathrm{d}\,E$, the vertical lines mark the integration boundaries.}
    \label{SI_fig:sensEQE}
\end{figure}

We quantify $\Voc$ losses by decomposing the total voltage loss into three contributions using a detailed-balance analysis based on $\EQE$ (Figure~\ref{SI_fig:sensEQE}). Specifically, we determine (i) the unavoidable radiative voltage loss, (ii) an additional radiative loss originating from the non-ideal absorption onset, and (iii) the non-radiative recombination loss. All quantities reported below were calculated consistently from $\EQE$.

We extract the optical bandgap $\Eg$ from the measured $\EQE$ following Rau et al,\cite{rau2017efficiency} i.e.\ from the derivative of the $\EQE$ spectrum with respect to photon energy. The derivative $\mathrm{d\,EQE}_{\mathrm{PV}}/\mathrm{d}\,E$ was treated as a distribution of ideal absorption onsets $P(E)$, and $\Eg$ was defined as the mean-peak energy \ref{SI_eq:meanpeak} of this distribution with the integration borders being set at half the maximum of the derivative (Figure~\ref{SI_fig:sensEQE}).
\begin{equation}
    E_g=\frac{\int_a^b E\,P(E)\mathrm{d}E}{\int_a^b P(E)\mathrm{d}E} \qquad \qquad a,b=\frac{1}{2}\mathrm{max}(|P(E)|)
    \label{SI_eq:meanpeak}
\end{equation}
Next, we calculated $\Voc$ in the Shockley--Queisser limit, $\Vocsq$, assuming an ideal step-function absorptance with a sharp onset at $\Eg$. The corresponding short-circuit current density and radiative dark saturation current density were obtained by integrating the AM1.5G solar photon flux ($\Phi_\mathrm{sun}$) and the 295\,K black-body photon flux ($\Phi_\mathrm{BB}$) above $\Eg$, respectively, yielding $\Vocsq$.\cite{yao2015quantifying}
\begin{equation}
\Vocsq = \frac{\kT}{e}
\ln\!\left(\frac{\jsc^{\mathrm{SQ}}}{J_{0}^{\mathrm{SQ}}}+1\right) = \frac{\kT}{e} \ln\!\left( \frac{e \int_{\Eg}^{\infty} \Phi_{\mathrm{sun}}(E)\,\mathrm{d}E}
{e \int_{\Eg}^{\infty} \Phi_{\mathrm{BB}}(E,T)\,\mathrm{d}E} + 1 \right) . 
\end{equation}

The radiative open-circuit voltage, $\Vocrad$, was calculated by replacing the step-function absorptance with the measured $\EQE(E)$. 
\begin{equation}
\Vocrad 
= \frac{\kT}{e}\ln\left(\frac{\jsc}{J_{0}^{\mathrm{rad}}}+1\right)
= \frac{\kT}{e}\ln\bl \frac{e \int_{0}^{\infty} \Phi_{\mathrm{sun}}(E)\,\EQE(E)\,\mathrm{d}E}{e \int_{0}^{\infty} \Phi_{\mathrm{BB}}(E,T)\,\EQE(E)\,\mathrm{d}E}+1 \br
\end{equation}

The open-circuit voltage losses were decomposed into three contributions,\cite{yao2015quantifying}
\begin{equation}
\Delta \Voc 
= \frac{\Eg}{e} - \Voc
= \underbrace{\left(\frac{\Eg}{e} - \Vocsq\right)}_{\dVocsq}
+ \underbrace{\left(\Vocsq - \Vocrad\right)}_{\dVocabs}
+ \underbrace{\left(\Vocrad - \Voc\right)}_{\dVocnr}.
\end{equation}
Here, $\dVocsq$ is the unavoidable radiative loss for an optical gap $\Eg$, $\dVocabs$ accounts for additional radiative losses associated with the non-ideal absorption onset, and $\dVocnr$ captures non-radiative recombination losses. The extracted values of $\Eg$, $\Vocsq$, $\Vocrad$, and the resulting loss terms are summarised in Table~\ref{SI_table:voc_losses}.

\begin{table}[h]
    \centering
    \caption{Calculated $\Voc$ losses for printed and spin-coated PM6:Y12 solar cells based on chloroform and o-xylene. Minor deviations of $\Delta \Voc$ arise from rounding.}
    \begin{tabular}{|c|cccc|ccc|}
        \hline
        Device type & $\Eg$\,[eV] & $\Vocsq$\,[V] & $\Vocrad$\,[V] & $\Voc$\,[V] & $\dVocsq$\,[V] & $\dVocabs$\,[V] & $\dVocnr$\,[V] \\
        \hline
        spin-coated, chloroform         & 1.412 & 1.152 & 1.082 & 0.834 & 0.260 & 0.070 & 0.248 \\
        spin-coated, o-xylene           & 1.412 & 1.152 & 1.084 & 0.831 & 0.260 & 0.068 & 0.253 \\
        gravure-printed, chloroform     & 1.447 & 1.185 & 1.094 & 0.801 & 0.262 & 0.091 & 0.293 \\
        gravure-printed, o-xylene       & 1.446 & 1.184 & 1.094 & 0.781 & 0.262 & 0.089 & 0.313 \\
        \hline
    \end{tabular}
    \label{SI_table:voc_losses}
\end{table}

\section{The recombination ideality factor}\label{SI_sec:nid}

\begin{figure}[h]
    \centering
    \includegraphics[width=0.45\linewidth]{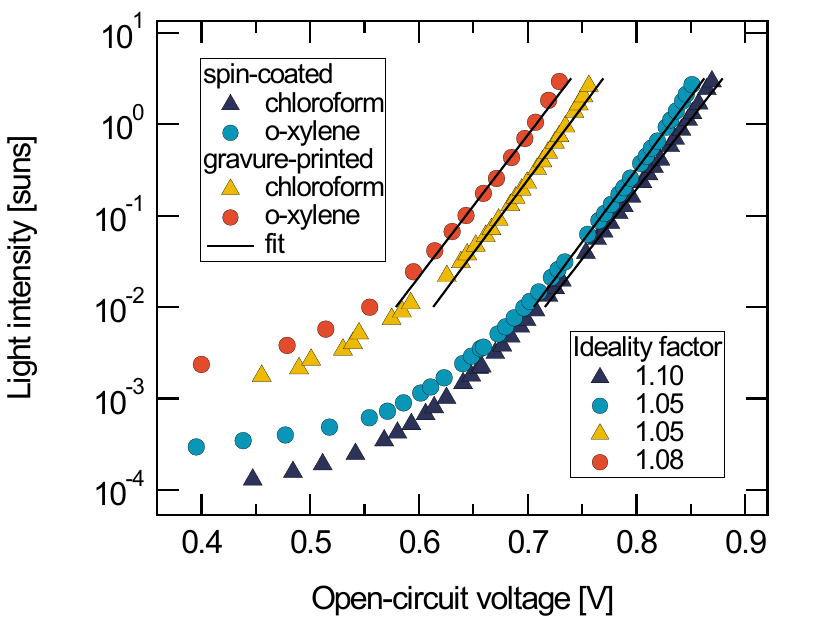}
    \caption{Recombination analysis of spin-coated and gravure-printed PM6:Y12 solar cells processed from chloroform and o-xylene. $\Voc$ as a function of light intensity with fits used to extract the recombination ideality factor $\nid$.}
    \label{SI_fig:I4-nid}
\end{figure}

The recombination ideality factor $\nid$ was extracted from light-intensity-dependent $\Voc$ measurements shown in Figure~\ref{SI_fig:I4-nid}, following the relation \cite{tvingstedt2016temperature}
\begin{equation}
\nid = \frac{e}{\kT} \bl \frac{\der\,\ln \Phi}{\der\,\Voc} \br^{-1} . 
\end{equation}
Here, $\Phi$ is the light intensity, $\kT$ is the thermal energy, and $e$ is the elementary charge. The value of $\nid$ was obtained from the slope of $\ln\Phi$ versus $\Voc$. The analysis was restricted to illumination ranges where the device response is not influenced by leakage currents at low light intensities or by contact-related limitations at high light intensities, ensuring that the extracted ideality factor reflects recombination processes only.

We must note that $\Voc$ values in Figure~\ref{SI_fig:I4-nid} for printed devices are much lower compared to Table~\ref{tab:PCE}. In the latter case, the JV curves were measured under 1\,sun equivalent illumination, with the full active area of 0.2\,cm$^2$ (defined as an overlap of silver and IMI) illuminated. In the former case, however, the sample was illuminated trough a small window of $\sim$0.05\,cm$^2$. This led to a mismatch between the active areas generating dark saturation current and photocurrent, and as the result, to overall lower $\Voc$. In short, $\Voc$ can be defined via the diode equation as
$$
\Voc \approx \frac{\nid\kT}{e}\,\ln\bl \frac{I_\mathrm{photo}}{I_0} \br
= \frac{\nid\kT}{e}\,\ln\bl \frac{J_\mathrm{photo} \cdot A_\mathrm{illu}}{J_0 \cdot A_\mathrm{total}} \br , 
$$
where $I_\mathrm{photo}$ denotes the current generated by light, $I_0$ is the current generated in the dark, $A_\mathrm{illu}$ is the illuminated area, and $A_\mathrm{total}$ is the total area contributing to $I_0$. To determine $\Voc$ correctly, one has to illuminate the whole active area. In this case, $A_\mathrm{illu} = A_\mathrm{total}$. We denote it $\Voc^\mathrm{true}$ to separate it from the one where the areas are not equal. The above equation can be expressed as a true $\Voc$ and an additional contribution coming from the difference of the areas:
$$
\Voc 
\approx \Voc^\mathrm{true} + \frac{\nid\kT}{e}\,\ln\bl \frac{A_\mathrm{illu}}{A_\mathrm{total}} \br , 
$$
In the case of printed solar cells, the area fraction was $\sim$0.25, and assuming $\nid = 1$ and $\kT/e = 0.026$\,V, the second term is $- 36$\,mV, which agrees well with the mismatch between the $\Voc$ measured through the aperture and the experimental value measured at the glovebox. 

It is important to note, however, that even if the $\Voc$ measured through the aperture is smaller, it does not affect the determination of the ideality factor. Since the procedure involves using the derivative, the area fraction just falls away as the derivative of a constant.

\section{Simulations and Machine Learning Methods}\label{SI_sec:ML}

To better interpret the experimental results, we developed a machine-learning framework to extract microscopic transport and recombination parameters from a single illuminated JV curve. The machine learning framework, learnt the inverse mapping between the JV curves and the electrical/material parameters that generated them. Training data was generated using the drift--diffusion simulator OghmaNano, which solves the coupled Poisson and charge-carrier continuity equations under steady-state illumination. The model explicitly includes energy-resolved Shockley-Read-Hall (SRH) trapping and recombination, allowing charge-carrier trapping and recombination via distributed trap states to be resolved in energy space. Parasitic series and shunt resistances are also included.

For each device architecture (printed and spin-coated), the full device geometry and optical stack were configured to match the corresponding experimental structure. The electrical parameters were then randomly varied within physically realistic bounds, including electron and hole mobilities, series and shunt resistances, trap-related free-to-trapped charge-carrier ratios, and SRH recombination rates at both transport layers. Parameters spanning multiple orders of magnitude were sampled logarithmically. Each simulation produced an illuminated JV curve sampled on a fixed voltage grid, resulting in paired data consisting of the JV characteristic and the corresponding vector of microscopic device parameters $\boldsymbol{P}$.

\subsection{Data preprocessing and representation}

The current density values were normalised to improve numerical stability during training. No handcrafted features were used, and the neural networks have been fed the discretised JV curves vectors. Targets were transformed when appropriate (e.g. they've been mapped to the logarithmic scale when strictly positive) to stabilize the loss landscape and prevent large-magnitude parameters to dominate over the smaller ones during the training process.

\subsection{Neural network architectures}

Two main neural network strategies were implemented: a standard point neural network (used as a reference baseline) and a differential objective method. Both architectures, at their core, are fully connected feed-forward neural networks operation on one-dimensional, discretised JV curves.

The baseline point network directly maps a JV curve $J\left(V\right)$ to a selected microscopic parameter:
$$\mathit{f}_{\theta}: J(V) \rightarrow p , $$
where $p \in \boldsymbol{P}$ can be any of the aforementioned target parameters, and $\theta$ represents the weights values of a specific neural network instance. The architecture consists of an input layer that receives the JV vector as input, then several hidden layers with ReLU activation, and then the output layer that returns the final parameter prediction. The loss function is the mean squared error (MSE) between the predicted and the true parameter values. Hence, this model provides a deterministic, single-point estimate for the selected parameter. By training different instances for each parameter we get a prediction for the full ensemble.

\subsection{Differential objective method}

Instead of learning the absolute mapping $J(V) \rightarrow p$, this method trains the neural network to predict parameter's differences between two devices:
$$\mathit{f}_\theta: \left(J_i(V), J_j(V)\right) \rightarrow \Delta p_{ij} , $$
where $\Delta p_{ij} = p_i - p_j$. This problem formulation offers two major advantages, the first one being a quadratic expansion of the original dataset: for a dataset of size $N$, all ordered pairs $\left(i, j\right)$ can be constructed, generating $N\left(N-1\right)$ training examples; since the dataset is constructed through expensive drift-diffusion simulations, this free data augmentation strategy enables huge scale-up in the applicability of neural networks. The second benefit is that predicting differences instead of the actual parameter values simplifies the regression tasks. In fact, learning differences reduces the range of the target space, while also focusing the neural network on relative variations. Moreover, these variations are more strictly related to the shape differences between the input JV curves.

\begin{figure}[ht]
    \centering
    \includegraphics[width=0.6\linewidth]{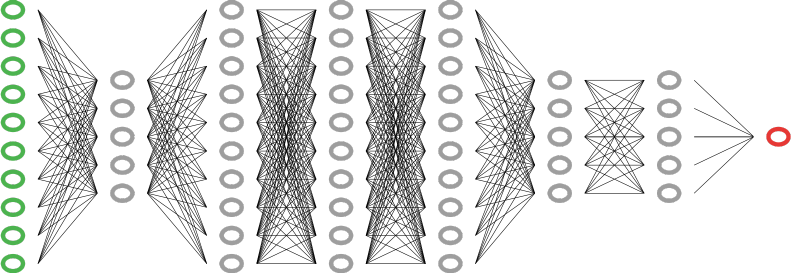}
    \caption{Schematic of the differential objective network. The two input JV curves are concatenated into a single vector and passed through a normalisation layer (grey rectangle), followed by three fully-connected hidden layers (grey nodes) of decreasing width. The single output node (red) predicts the parameter difference $\Delta p_{ij}$ between the two input devices.}
    \label{SI_fig:ML_diff_method}
\end{figure}

This architecture is illustrated schematically in Figure \ref{SI_fig:ML_diff_method}, showing the normalised concatenated input vector passing through successive fully-connected hidden layers to a single output node predicting $\Delta p_{ij}$.

\subsection{Differential objective method with residual blocks}

To further enhance the differential method, a second variant was developed that introduces two key architectural changes. First, residual blocks\cite{he2015deepresiduallearningimage} are used throughout in place of standard dense layers, mitigating the vanishing gradient problem in deeper networks and stabilising training. Mathematically, a residual block is described as:
$$x_{t+1} = x_t + F(x_t) , $$
where $x_t$ is the input of the $t$-th network layer, $F(x_t)$ is the transformation learned by the same layer and $x_{t+1}$ is the input of the following (i.e. $t+1$-th) layer; in a nutshell, we are feeding to a following layer the output of the previous one summed to it input. This residual learning is known to stabilize the training process, enabling deeper architectures without performance degradation.\cite{he2015deepresiduallearningimage}

Second, rather than concatenating the two input JV curves, this variant processes them through two completely independent and parallel branches. Each branch begins with a dense projection layer followed by a stack of residual blocks, learning a latent embedding of its input curve. The two embeddings are then merged via element-wise subtraction:
$$x = \mathit{f}\left(J_i(V)\right) - \mathit{f}\left(J_i(V)\right) . $$
This subtraction is physically motivated: if both branches learn the same latent representation of device behaviour, their difference encodes the relative change in underlying material parameters. The subtracted representation is then passed through a further stack of layers before the final output layer produces the predicted parameter difference $\Delta p_{ij}$.

\begin{figure}[ht]
    \centering
    \includegraphics[width=0.9\linewidth]{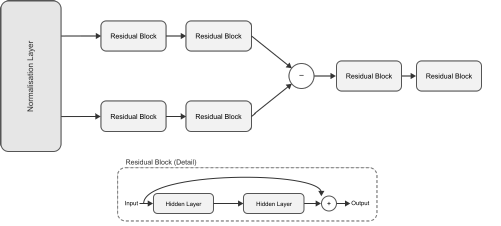}
    \caption{Schematic of the differential objective network with residual blocks. The two input JV curves enter through a shared normalisation layer and are then processed by two independent parallel branches, each consisting of two residual blocks. The learned embeddings from the two branches are merged via element-wise subtraction ($-$ node), and the result is passed through two further residual blocks to produce the final parameter difference prediction $\Delta p_{ij}$. The lower square details the internal structure of a residual block: the input passes through two hidden layers while also being routed via a skip connection, with the two paths summed at a addition node ($+$) before the output, implementing the transformation $x_{t+1} = x_t + F(x_t)$.}
    \label{SI_fig:ML_res_method}
\end{figure}

This architecture is illustrated schematically in Figure \ref{SI_fig:ML_res_method}, where the two input JV curves are processed by independent parallel branches, each consisting of two residual blocks, before being merged by the subtraction node. The resulting representation is then passed through further residual blocks towards the output. The inset details the internal structure of each residual block, showing the skip connection that adds the block input directly to its output.

\subsection{Training and inference procedures}

Training was performed using mini-batch gradient descent with adaptive optimisation, in particular using the Adam algorithm.\cite{kingma2017adam} The dataset was split into training and validation subsets to monitor over-fitting and general training evolution. Moreover, early stopping (based on validation loss) was employed to stop training before time is the performance stopped improving halfway.

During inference, for each target device, its JV curve is paired with multiple simulated reference curves. For each pair, the neural network predicts the parameter difference, and then the absolute parameter estimation is reconstructed by adding the predicted difference to the known simulated references. By aggregating all these predictions, the method then returns a distribution of parameter estimates. The final prediction is the mean of the aforementioned distribution, while the standard deviation quantifies the predictive spread. It is important to highlight that this uncertainty does not rely on assuming a Gaussian prior for the distribution, since it emerges directly from the ensemble of differential predictions.

\subsection{Results}

ML predictions were obtained for all the four devices analysed in the paper. The results discussed below have been obtained using the Differential Objective method with residual block discussed before. For each parameter, the reported value corresponds to the mean of the empirical prediction distribution. Predictive reliability is estimated using the validation MAPE.

Table~\ref{tab:ml_core_parameters} shows the predicted harmonic average mobilities and recombination lifetimes. The machine-learning inversion first indicated that the intrinsic free-carrier mobilities in the spin-coated devices were higher than those in the printed counterparts, in some cases approaching an order-of-magnitude difference. However, these values represent trap-free transport parameters and therefore exceed the effective mobilities typically inferred from experiment. Once trap-mediated transport was explicitly accounted for, the resulting harmonic mobilities at $\jsc$ converged to the $10^{-8}\,\mathrm{m^2\,V^{-1}\,s^{-1}}$ range for both printed and spin-coated devices, yielding values consistent with experimentally accessible transport scales under operating conditions. The recombination time constants extracted at $\Voc$ were of the same order of magnitude for printed and spin-coated devices ($10^{-6}\mathrm{s}$). While minor variations were observed, no statistically significant differences emerged within the uncertainty of the inversion.
The training MAPE of the of the values previously discussed are rather low, all between 4\% and 12\%, which indicates a high predicting accuracy of the neural networks with regard to the training datasets.

\begin{figure}[ht]
    \centering
    \includegraphics[width=0.8\linewidth]{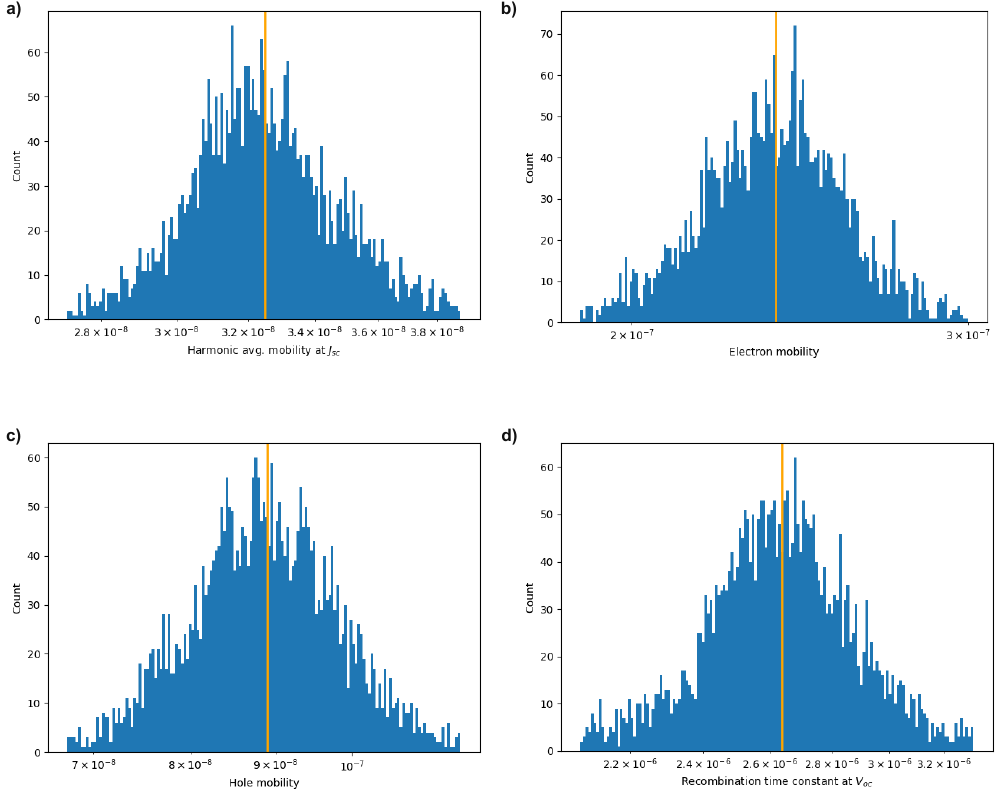}
    \caption{Predictions distributions for the parameters presented in Table~\ref{tab:ml_core_parameters}. The narrow spread of the histograms shows high reliability in the final values predicted for the experimental cells. Outliers were filtered using IQR.}
    \label{SI_fig:ml_histograms_good}
\end{figure}

\begin{figure}[ht]
    \centering
    \includegraphics[width=0.8\linewidth]{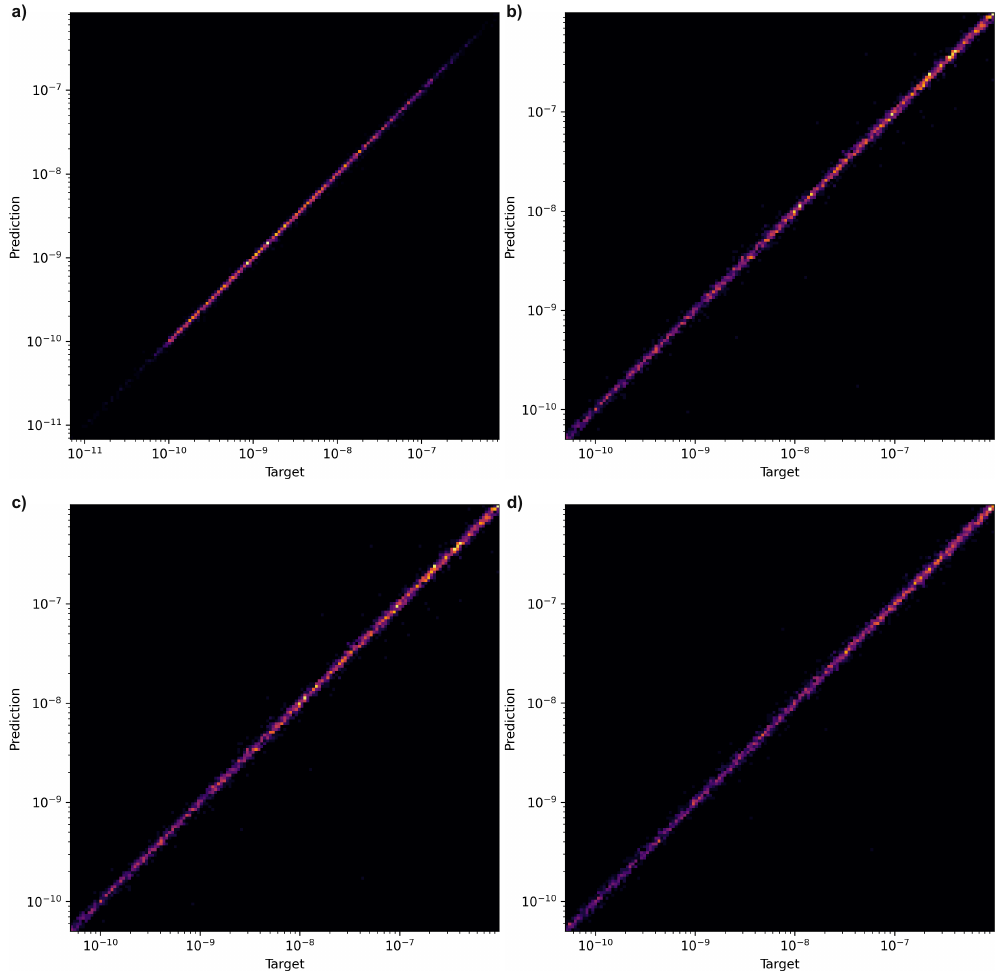}
    \caption{Confusion matrices for the parameters presented in table \ref{tab:ml_core_parameters}. Thin lines mean that the values are clustered around the $y = x$ line, indicating strong similarity between predicted and true values in the held-out validation set.}
    \label{SI_fig:ml_conf_matrices_good}
\end{figure}

Figures~\ref{SI_fig:ml_histograms_good} and \ref{SI_fig:ml_conf_matrices_good} show the resulting distribution and confusion matrices for the reported mobilities and the recombination time constant at $\Voc$.

\begin{table*}[t]
\centering
\caption{Machine-learning extracted core device parameters for gravure-printed and spin-coated PM6:Y12 solar cells. Values correspond to the filtered mean of the inversion ensemble.}
\label{tab:ml_core_parameters}
\renewcommand{\arraystretch}{1.2}
\begin{tabular}{lcccc}
\hline
Parameter & Printed CF & Spin CF & Printed OXY & Spin OXY \\
\hline
Harmonic mobility at $\jsc$ ($\mathrm{m^2\,V^{-1}\,s^{-1}}$) 
& $3.25\times10^{-8}$ & $2.35\times10^{-8}$ & $2.35\times10^{-8}$ & $3.31\times10^{-8}$ \\

Electron mobility ($\mathrm{m^2\,V^{-1}\,s^{-1}}$)
& $2.38\times10^{-7}$ & $2.33\times10^{-7}$ & $2.70\times10^{-7}$ & $3.24\times10^{-6}$ \\

Hole mobility ($\mathrm{m^2\,V^{-1}\,s^{-1}}$)
& $8.90\times10^{-8}$ & $1.90\times10^{-7}$ & $1.41\times10^{-7}$ & $3.67\times10^{-7}$ \\

Recombination lifetime at $\Voc$ (s)
& $2.64\times10^{-6}$ & $2.28\times10^{-6}$ & $3.25\times10^{-6}$ & $3.09\times10^{-6}$ \\

\hline
\end{tabular}
\end{table*}

\clearpage
\section{Determination of transport resistance losses}\label{SI_sec:I4-JV}

\subsection{Transport resistance}

To quantify $\FF$ losses, we analyse the JV characteristics using the transport-resistance framework.\cite{wurfel2015impact,neher2016new,saladina2025transport} In this approach, the applied voltage is separated into a component associated with the quasi-Fermi level splitting and an additional voltage drop arising from charge transport.

We first consider the idealised case without transport resistance, where the externally applied voltage equals the implied voltage $\Vimp$ associated with the quasi-Fermi level splitting. In this limit, the JV curve is governed solely by recombination losses, and the ideality factor reflects recombination only. The corresponding pseudo-JV curve is given by
\begin{equation}\label{SI_eq:jVimp}
    J(\Vimp)
    = \jgen \bL \exp\!\left( \frac{e\Vimp-e\Voc}{\nid\kT} \right) -1 \bR , 
\end{equation}
where $\jgen$ is the generation current density. For each device, the pseudo-JV curve was calculated using the ideality factors determined from the suns--$\Voc$ measurements and $\jgen$ estimated from the reverse-bias current at $-0.5$\,V.

\begin{figure}[b]
    \centering
    \includegraphics[width=0.9\linewidth]{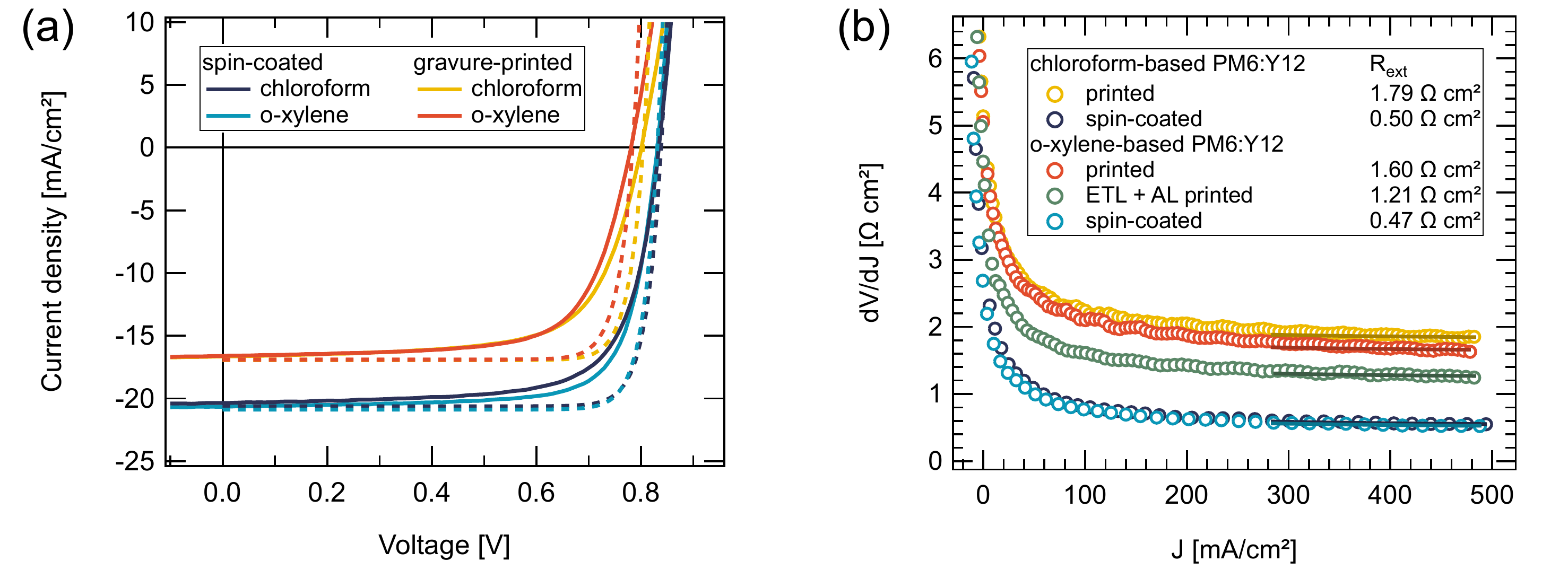}
    \caption{Transport resistance analysis of spin-coated and gravure-printed PM6:Y12 solar cells processed from chloroform and o-xylene. (a) Measured $J(\Vext)$ (solid lines) under 1\,sun equivalent illumination and the corresponding pseudo-JV (dashed lines) used for the determination of transport-related fill factor losses. (b) External series resistance $\Rext$ is determined from the fit of $\der V/ \der J$ in the forward bias region using Eq.~\eqref{SI_eq:Rext}.}
    \label{SI_fig:I4-JV}
\end{figure}

In real devices, charge transport introduces an additional voltage drop, leading to a shallower JV slope and an effective ideality factor that contains contributions from both recombination and transport. This behaviour is described by
\begin{equation}\label{SI_eq:jVext}
    J(\Vext)
    = \jgen \bL \exp\!\left( \frac{e\Vext-e\Voc}{(\nid+\beta)\kT} \right) -1 \bR , 
\end{equation}
where $\Vext$ is the externally applied voltage and $\beta$ quantifies the strength of transport-induced voltage losses. The pseudo-JV curves without transport resistance and the experimentally measured JV curves are compared in Figure~\ref{SI_fig:I4-JV}(a).

In general, $\beta$ depends on the operating point of the device. A special case is the open-circuit condition, where $\beta$ reduces to the figure of merit $\alpha$, originally introduced by Neher et al.\cite{neher2016new} The parameter $\alpha$ is related to the effective conductivity at open circuit, $\sigmaoc$, via \cite{saladina2025transport}
\begin{equation}\label{SI_eq:alpha}
    \alpha = \frac{eL}{\kT}\,\frac{\jgen}{\sigmaoc} , 
\end{equation}
where $L$ is the active-layer thickness. The open-circuit conductivity was extracted from the slope of the illuminated JV curve at $J=0$ according to \cite{saladina2025transport}
\begin{equation}\label{SI_eq:sigmaoc}
    \frac{L}{\sigmaoc}
    = \left. \left( \frac{\mathrm{d}\Vext}{\mathrm{d}J}
        - \frac{\nid\kT}{e\jgen} -\Rext \right) \right|_{J=0} . 
\end{equation}

We determined the external series resistance $\Rext$ by fitting $\der V/\der J$ at high forward bias. There, the transport resistance is assumed to be much smaller than $\Rext$. Then 
\begin{equation}\label{SI_eq:Rext}
    \frac{dV\bl J \br}{d J} \approx \frac{\nid\kT}{e} \cdot \frac{1}{J +\jgen} + \Rext . 
\end{equation}
The fit along with the values for the gravure-printed and spin-coated devices as well as half-printed stacks is shown in Figure~\ref{SI_fig:I4-JV}(b). 

While $\alpha$ provides a convenient measure of transport losses, it is defined at open circuit and therefore tends to overestimate the $\FF$. To obtain a more accurate description of transport losses at the maximum power point, we determine the effective transport parameter $\beta_\mathrm{MPP}$ using an iterative scheme,\cite{saladina2025transport}
\begin{equation}\label{SI_eq:betampp}
    \beta_\mathrm{MPP}
    = \alpha \cdot \frac
        { \uoc\,(\uoc+1)^{\frac{\nid}{n_\sigma}-1} }{ \ln(\uoc+1) },
    \qquad
    \uoc 
    = \frac{e\Voc}{(\nid+\beta_\mathrm{MPP})\kT},
\end{equation}
which self-consistently accounts for the voltage dependence of transport losses. In the above equation, $\nsig$ stands for the transport ideality factor, expressing the voltage dependence of charge transport, similarly as the recombination ideality factor $\nid$ does for recombination. The ratio of the ideality factors was assumed to be equal to 0.5. The resulting $\beta_\mathrm{MPP}$ values are used to quantify transport-induced $\FF$ losses in the main text.

\subsection{Predicting the fill factor}

Beyond quantifying $\FF$ losses, the transport-resistance framework allows prediction of the $\FF$ from physical parameters. Three equations are required. Equation~\eqref{SI_eq:alpha} defines the transport figure of merit $\alpha$ at $\Voc$. Equation~\eqref{SI_eq:betampp} uses $\alpha$ to iteratively determine the transport parameter $\beta$ at the maximum power point. Finally, the Green equation predicts the $\FF$.\cite{green_accuracy_1982} Because the $\FF$ is affected by both recombination and transport resistance, the recombination ideality factor $\nid$ is replaced by the apparent ideality factor $(\nid + \beta_\mathrm{MPP})$:\cite{saladina2025transport}
\begin{equation}\label{SI_eq:FF_green}
\FF 
= \frac{\uoc - \ln\!\left(\uoc + 0.72\right)}{\uoc + 1} ,
\qquad
\uoc 
= \frac{e\Voc}{\bl\nid + \beta_\mathrm{MPP}\br\kT} .
\end{equation}

Let us calculate the $\FF$ of the printed PM6:Y12 solar cell for the case, where its layer stack is optimised, i.e.\ the photocurrent and charge-carrier density is approximately equal to those of the spin-coated device. In this case, the only variable is mobility (IMPS mobilities are shown in Figure~\ref{fig:voc_ff_losses}(d)). Since $\sigma = en\mu$, the figure of merit $\alpha$ (Eq.~\eqref{SI_eq:alpha}) for the printed device would increase by approximately one order of magnitude compared to the printed one:
$$\alpha_\mathrm{print} = \alpha_\mathrm{spin} \cdot \frac{\mu_\mathrm{spin}}{\mu_\mathrm{print}} , $$
and thus $\alpha_\mathrm{print} = 5.1$.

Using the iterative scheme of Eq.~\eqref{SI_eq:betampp}, we obtain $\beta_{\mathrm{MPP,print}} = 5.8$. This corresponds to the expected transport loss at the maximum power point. We assume $\Voc = 831$\,mV, identical to the spin-coated device, and set $\nid/\nsig = 0.5$ for simplicity. The iteration converges rapidly, with changes below the third decimal after three steps. Substituting the values into Eq.~\eqref{SI_eq:FF_green} yields a predicted $\FF$ of 52.75\,\%. The above calculation excludes external series resistance losses since $\alpha$ is corrected for $\Rext$. 

To validate the approach, we apply the same calculation to the spin-coated reference device, for which the $\FF$ is known. We obtain $\alpha_\mathrm{spin} = 0.7$, $\beta_{\mathrm{MPP,spin}} = 0.9$, and a predicted $\FF$ of 78.01\,\%, in close agreement with the value of 76.76\,\% for o-xylene-based spin-coated solar cell when the fill factor is corrected for external series resistance. This confirms that using $\beta_{\mathrm{MPP}}$ provides a reliable prediction of the $\FF$.

\clearpage
\section{Biased external quantum efficiency}

\begin{figure}[h]
    \centering
    \includegraphics[width=0.9\linewidth]{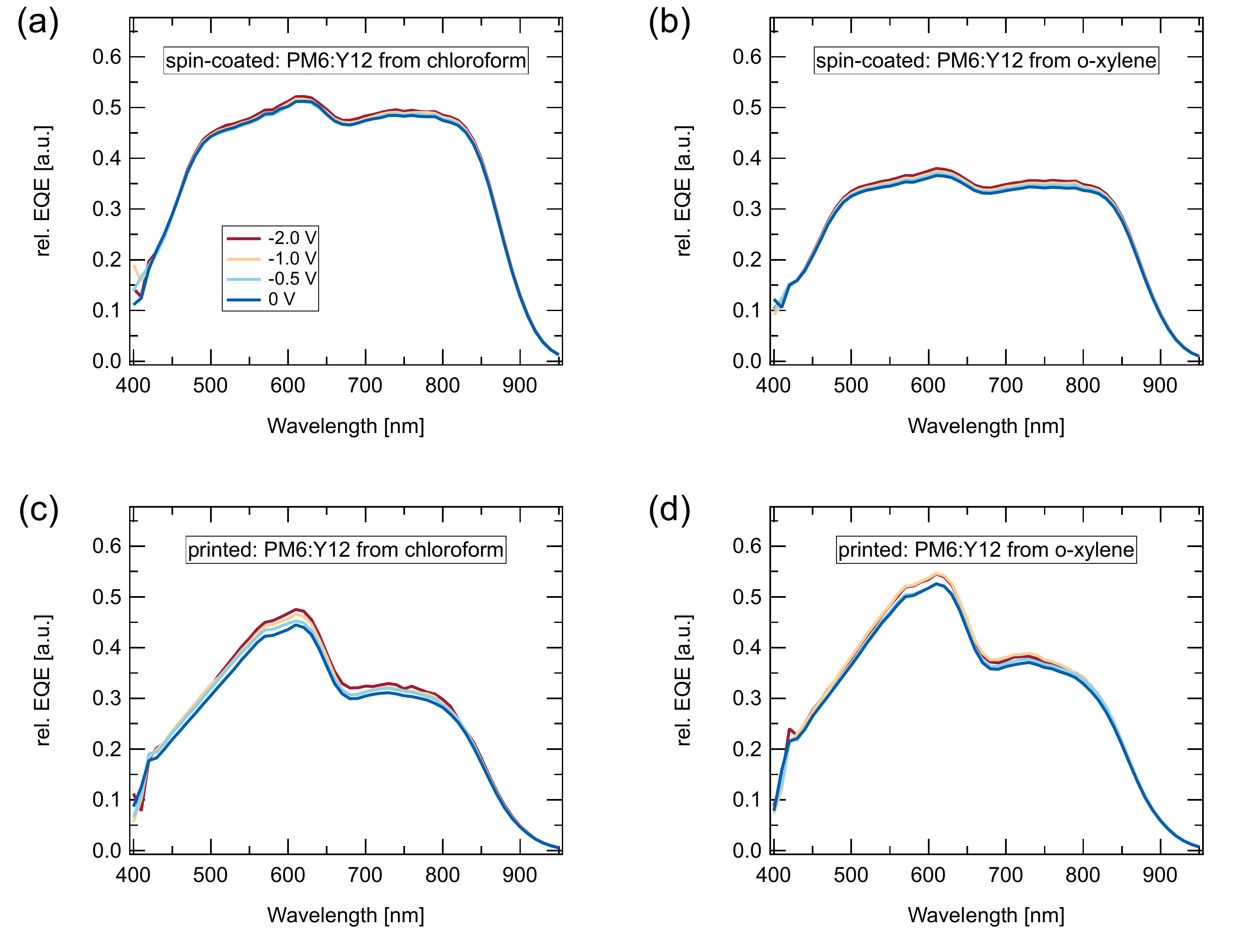}
    \caption{Relative EQE measured under different applied bias voltages for (a, b) spin-coated and (c, d) printed PM6:Y12 solar cells processed from (a, c) chloroform and (b, d) o-xylene.}
    \label{SI_fig:biasedEQE}
\end{figure}

Figure~\ref{SI_fig:biasedEQE} shows the relative EQE measured under different applied bias voltages for spin-coated and printed PM6:Y12 solar cells processed from chloroform and o-xylene. For all devices, the spectral shape of the EQE remains largely unchanged with bias, indicating that the underlying photogeneration and charge-transfer processes are not strongly field dependent. At the same time, a gradual increase in the EQE magnitude is observed with increasing reverse bias, reflecting improved charge extraction.

\clearpage
\section{Charge-carrier mobility from intensity-modulated photocurrent spectroscopy}\label{SI_sec:IMPS}

From the IMPS measurements, the real and imaginary components of the photocurrent response, $\mathrm{Re}(\text{IMPS})$ and $\mathrm{Im}(\text{IMPS})$, were obtained as a function of angular frequency $\omega = 2\pi f$. The frequency $\omega_{\mathrm{peak}}$ corresponds to the maximum of $\mathrm{Im}(\text{IMPS})$, which is determined by Gaussian fit. $\omega_{\mathrm{peak}}$ provides a measure of the charge-carrier transition time $\tau_{\mathrm{tr}}$ from generation to their arrival at either electrode under short-circuit conditions, $\tau_{\mathrm{tr}} = \pi / (2\omega_{\mathrm{peak}})$.\cite{nojima2019modulated}

Assuming only drift current, uniform photogeneration and electric field across the active layer and position-independent mobility, the effective charge-carrier mobility $\mu$ can be estimated as $\mu = L / (2 \tau_{\mathrm{tr}} F)$, where $L$ is the active layer thickness and $F$ is the internal electric field. Under short-circuit conditions in IMPS measurements, the electric field is approximated by $F = V_\mathrm{bi} / L$, yielding 
\begin{equation}
    \mu = \frac{\omega_{\mathrm{peak}} L^2}{\pi V_{\mathrm{bi}}}.
\end{equation}

\begin{figure}[h]
    \centering
    \includegraphics[width=0.9\linewidth]{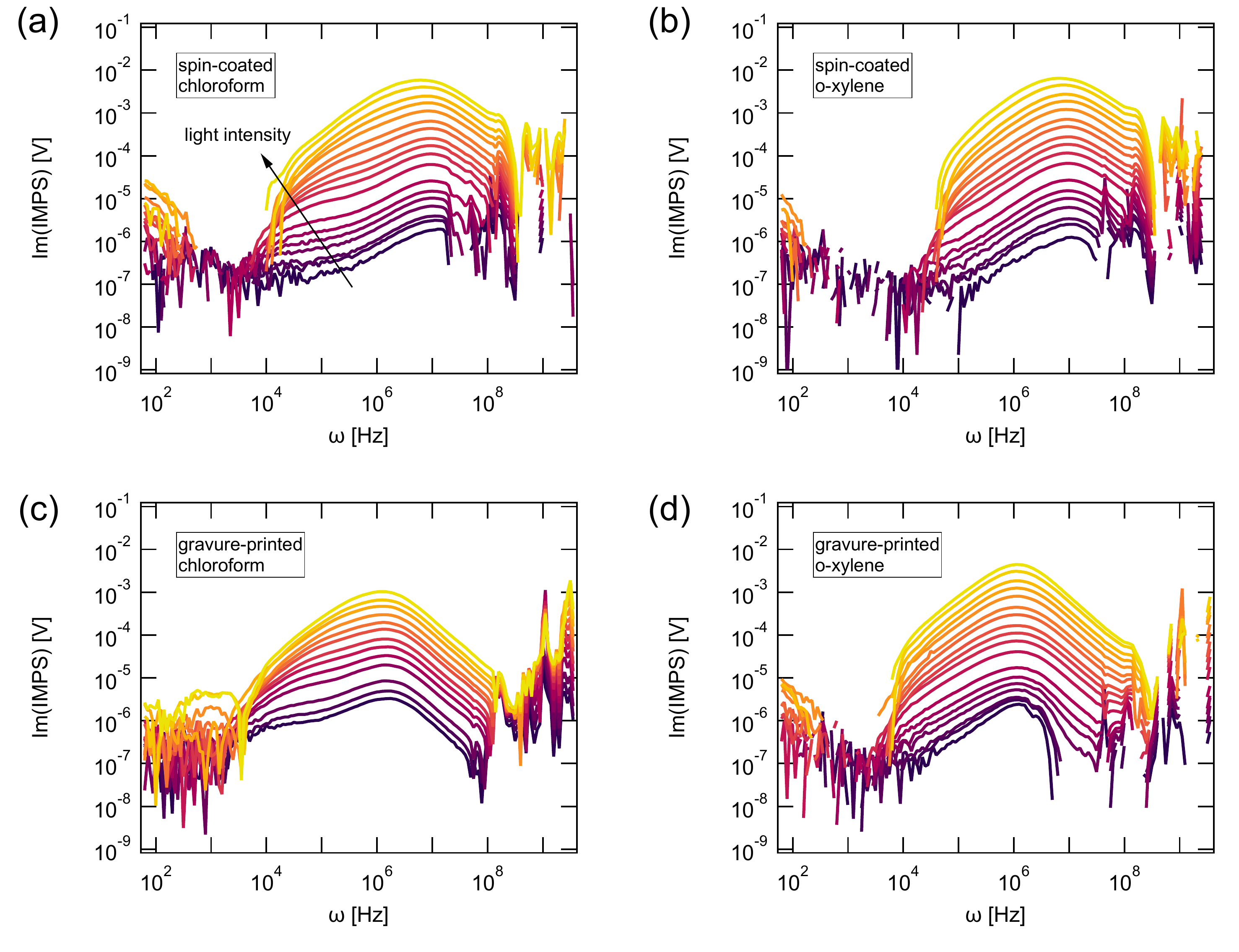}
    \caption{$\mathrm{Im}(\text{IMPS})$ as a function of $\omega$ for (a, b) spin-coated and (c, d) printed PM6:Y12 devices based on (a, c) chloroform and (b, d) o-xylene.}
    \label{SI_fig:IMPS_raw}
\end{figure}

The measured $\mathrm{Im}(\text{IMPS})$ is shown in Figure~\ref{SI_fig:IMPS_raw} and the resulting mobility values in Figure~\ref{fig:voc_ff_losses}(d). The built-in voltage was assumed to be $1.15$\,V for all devices. 

\end{document}